\documentclass[acmsmall,screen]{acmart}

\usepackage{pifont}
\usepackage{tikz}
\usepackage{mdframed}
\usepackage{multirow}
\usepackage[normalem]{ulem}

\usepackage{xcolor}

\usepackage{booktabs}      
\usepackage{array}         
\usepackage{threeparttable} 

\usepackage{longtable}     
\usepackage{ltxtable}      
\usepackage{multirow}      
\usepackage{makecell}      
\usepackage[commandnameprefix=always]{changes}  
\definechangesauthor[name={yulei}, color=red]{author1}

\AtBeginDocument{%
  }

\setcopyright{acmlicensed}
\copyrightyear{2024}
\acmYear{2024}
\acmDOI{XXXXXXX.XXXXXXX}

\acmConference[Conference acronym 'XX]{Make sure to enter the correct
  conference title from your rights confirmation emai}{June 03--05,
  2024}{Woodstock, NY}
\acmISBN{978-1-4503-XXXX-X/18/06}

\begin{document}


\title[Towards Secure and Explainable Smart Contract Generation with Security-Aware Group Relative Policy Optimization]{Towards Secure and Explainable Smart Contract Generation with Security-Aware Group Relative Policy Optimization}

\author{Lei Yu}
\authornote{Affiliated with University of Chinese Academy of Sciences, Beijing, China.}
\orcid{0000-0003-3134-3746}
\affiliation{
  \institution{Institute of Software, Chinese Academy of Sciences}
  \city{Beijing}
  \country{China}
}
\email{yulei2022@iscas.ac.cn}

\author{Jingyuan Zhang}
\authornotemark[1]
\orcid{0000-0001-5475-3815}
\affiliation{
  \institution{Institute of Software, Chinese Academy of Sciences}
  \city{Beijing}
  \country{China}
}
\email{zhangjingyuan2023@iscas.ac.cn}

\author{Xin Wang}
\authornotemark[1]
\orcid{0009-0005-5391-8821}
\affiliation{
  \institution{Institute of Software, Chinese Academy of Sciences}
  \city{Beijing}
  \country{China}
}
\email{wangxin@iscas.ac.cn}

\author{Jiajia Ma}
\orcid{0000-0002-6028-4186}
\affiliation{
  \institution{Institute of Software, Chinese Academy of Sciences}
  \city{Beijing}
  \country{China}
}
\email{majiajia@iscas.ac.cn}

\author{Li Yang}
\authornote{Li Yang and Fengjun Zhang are the corresponding authors.}
\orcid{0000-0001-8364-6525}
\affiliation{
  \institution{Institute of Software, Chinese Academy of Sciences}
  \city{Beijing}
  \country{China}
}
\email{yangli2017@iscas.ac.cn}

\author{Fengjun Zhang}
\authornotemark[2]

\orcid{0000-0002-3830-8786}
\affiliation{
  \institution{Institute of Software, Chinese Academy of Sciences}
  \city{Beijing}
  \country{China}
}
\email{fengjun@iscas.ac.cn}

\renewcommand{\shortauthors}{Yu et al.}

\begin{abstract}
Smart contracts automate the management of high-value assets, where vulnerabilities can lead to catastrophic financial losses. In the task of automated smart contract generation using Large Language Models (LLMs), this challenge is amplified by two interconnected failures: first, they operate as unauditable "black boxes" by failing to produce a transparent reasoning process, and second, as a consequence, they generate code riddled with critical security vulnerabilities. To address both issues, we propose SmartCoder-R1 based on Qwen2.5-Coder-7B, a novel framework for secure and explainable smart contract generation. It begins with Continual Pre-training (CPT) to specialize the base model on the nuances of smart contract code. To construct the data for subsequent stages, we first prompt the DeepSeek model to generate reasoning-and-code samples from verified on-chain contracts, followed by a rigorous validation process where each sample is manually reviewed by security experts for compilability, functionality, security, and reasoning completeness. Based on this, we then apply Long Chain-of-Thought Supervised Fine-Tuning (L-CoT SFT) on 7,998 of these expert-validated samples to train the model to emulate human security analysis. Finally, to directly mitigate vulnerabilities, we employ Security-Aware Group Relative Policy Optimization (S-GRPO), a reinforcement learning phase that refines the generation policy using 1,691 samples by optimizing a weighted reward signal for compilation success, security compliance, and format correctness. Evaluated against 17 state-of-the-art baselines on a challenging benchmark of 756 real-world functions from 289 deployed contracts, SmartCoder-R1 establishes a new state of the art by achieving top performance across five key metrics: a ComPass of 87.70\%, a VulRate of 8.60\%, a SafeAval of 80.16\%, a FuncRate of 53.84\%, and a FullRate of 50.53\%. This FullRate marks a 45.79\% relative improvement over the strongest baseline, DeepSeek-R1. Crucially, its generated reasoning also excels in human evaluations, achieving high-quality ratings for Functionality (82.7\%), Security (85.3\%), and Clarity (90.7\%).

\end{abstract}


\maketitle

\section{Introduction}

Blockchain technology has been rapidly adopted across various domains due to its decentralized architecture \cite{swan2015blockchain}. This innovative technology enables the creation of secure, distributed digital ledgers for recording transactions \cite{hewa2021survey}. By utilizing advanced cryptographic methods, blockchain ensures the integrity and verification of each transaction \cite{wood2014ethereum, yu2023money}. Within this ecosystem, smart contracts function as self-executing programs on the blockchain, automating the management of digital assets such as cryptocurrencies. These contracts are activated when specific conditions are met and, once deployed, become permanent components of the blockchain \cite{zou2019smart}. However, the immutability and inherent complexity of smart contracts pose significant security challenges \cite{zou2019smart}. The well-known DAO incident \cite{dhillon2017dao,mehar2019understanding} serves as a cautionary example, demonstrating the potential severity of such vulnerabilities. This security breach resulted in the illegal transfer of \$60 million worth of Ethereum, causing widespread impact on the blockchain community \cite{alharby2017blockchain, hegedHus2018towards}.

The primary challenge is that existing Code LLMs often lack a deep, contextual understanding of security principles, leading to two interconnected failures. First, they fail to produce an explicit reasoning process, operating as “black boxes” that prevent developers from auditing the security logic. Second, this shallow understanding results in code with critical vulnerabilities. For instance, as illustrated in our motivating example (Fig. \ref{tab:motivation}), a powerful model like \texttt{Qwen2.5-Coder-7B-Instruct}, when tasked to implement a \texttt{withdraw} function that prioritizes a developer fee, produces a flawed implementation. It correctly transfers the fee but then transfers the original, full balance to the recipient without deducting the fee, creating a severe “double payment” vulnerability. This error vividly demonstrates that even top-tier models cannot be trusted for secure contract generation without a more structured reasoning capability. This is not an isolated issue; empirical data from our broader experiments further substantiates that reasoning-focused models consistently outperform their standard counterparts. For instance, the reasoning-based \texttt{QwQ-32B} achieves a FullRate of 23.94\%, surpassing the 20.50\% achieved by the standard \texttt{Qwen2.5-32B-Instruct}. This pattern highlights that models equipped with explicit reasoning capabilities are better suited for the security-critical domain of smart contract generation.

Previous research has attempted to mitigate these risks from two main directions. The first is post-hoc vulnerability detection, which audits the code after it has been generated. Although recent work has leveraged Large Language Models (LLMs) for automated auditing with tools like GPTScan \cite{sun2024gptscan} and iAudit \cite{ma2024combining}, this paradigm has a fundamental weakness. It decouples code generation from security verification, forcing developers to first receive potentially black-box-generated code and then use a separate tool to find its flaws. This not only disrupts the development workflow but can also lead to complex and inefficient fixes, especially when the auditing and generation models are inconsistent (a risk highlighted by iAudit's separated architecture \cite{ma2024combining}). This approach is inherently reactive, aiming to 'patch insecurity' rather than proactively 'build in security'. The second direction focuses on proactive secure code generation. For instance, Storhaug et al. \cite{storhaug2023efficient} proposed a method based on vulnerability tagging and constrained decoding to proactively avoid known insecure patterns during generation, while CodeBC \cite{wang2025codebc} employs a three-stage fine-tuning strategy to enhance the security and practicality of the generated code. However, these approaches still often lack the explicit, multi-step reasoning required to handle complex business logic securely and fail to produce auditable thought processes. This leaves a critical gap for a model that not only generates secure code but also transparently explains how it achieved that security.

To bridge these gaps, we propose SmartCoder-R1 based on Qwen2.5-Coder-7B, a framework designed to generate smart contracts that are not only secure and functional but also inherently explainable by tackling the dual challenges of opaque reasoning and code vulnerability. Our approach is built on a meticulously designed three-stage training pipeline that ensures a logical closed loop from problem to solution. First, to build a foundational understanding of Solidity, Continual Pre-training (CPT) is performed on a comprehensive corpus of 286,397 instances. Second, to address the lack of transparency, we employ Long Chain-of-Thought Supervised Fine-Tuning (SFT) on 7,998 expert-validated samples. This stage explicitly teaches the model to emulate a security expert's thought process, generating a step-by-step reasoning chain (\texttt{<think>...</think>}) before writing the final code (\texttt{<answer>...</answer>}). Finally, to directly minimize vulnerabilities, we apply Security-Aware Group Relative Policy Optimization (S-GRPO) using 1,691 samples. This stage operates within a reinforcement learning paradigm, aligning the model (the policy network $\pi_\theta$) with a programmatic reward function $R$. For each input, the policy network generates a group of candidate outputs. The reward function $R$, a weighted sum of binary scores for compilation success ($R_{\text{compile}}$), security compliance ($R_{\text{security}}$), and format correctness ($R_{\text{format}}$), then evaluates each output. The core of S-GRPO lies in calculating a normalized advantage for each candidate relative to the group's average performance and using this signal to update the model's parameters $\theta$ via a policy gradient update. By maximizing the expected reward, this method directly steers the model's generation distribution towards verifiably secure, compilable, and well-structured code, thereby internalizing complex security constraints into the model's generation policy.

To validate our approach, we conducted extensive experiments on a challenging benchmark comprising 756 real-world functions sourced from 289 deployed contracts, covering a wide range of complexities from simple utilities to intricate business logic. The results confirm the superiority of SmartCoder-R1. It establishes a new state of the art, achieving top performance across five key metrics: a ComPass (proportion of successfully compiled code) of 87.70\%, a VulRate (proportion of compilable code with vulnerabilities; lower is better) of 8.60\%, and a SafeAval (proportion of all generated code that is both compilable and secure) of 80.16\%. Consequently, our model achieves a FuncRate (proportion of all generated code that is both compilable and functionally correct) of 53.84\%, culminating in a final FullRate (the overall proportion of code that is simultaneously compilable, secure, and functionally correct) of 50.53\%. This FullRate represents a substantial absolute improvement of 15.87 percentage points and a 45.79\% relative improvement over the strongest reasoning-based baseline, DeepSeek-R1, which scored 34.66\%. Furthermore, human evaluation of the reasoning process confirms its superiority in explainability. SmartCoder-R1's reasoning achieved high-quality ratings across all key dimensions of Functionality (82.7\%), Security (85.3\%), and Clarity (90.7\%), comprehensively outperforming baselines and proving the reliability of its thought process. 

The main contributions of this paper are as follows:
\begin{itemize}
\item We propose SmartCoder-R1, the first framework to systematically integrate Continual Pre-Training (CPT), Long Chain-of-Thought Supervised Fine-Tuning (L-CoT SFT), and Security-Aware Group Relative Policy Optimization (S-GRPO) for secure and explainable smart contract generation.
\item We construct and release novel, high-quality datasets with expert-validated reasoning chains, comprising 7,998 samples for SFT and 1,691 samples for S-GRPO, to foster community research in explainable smart contract generation.
\item We demonstrate that SmartCoder-R1 establishes a new state of the art, achieving a FullRate of 50.53\% and outperforming the strongest baseline by a relative margin of 45.79\%, proving its superior ability to generate verifiably secure and functional smart contract.
\end{itemize}

\section{Background and Motivation}

\subsection{Problem Statement}

Given a smart contract context $C$ and a functional requirement description $R$, the goal of smart contract code generation is to generate a function implementation $F$ such that $F$ satisfies the following constraints: \textbf{Functional Correctness}: $F$ must correctly implement all functional requirements specified in $R$. This includes proper handling of input parameters, correct state variable modifications, appropriate return values, and adherence to the specified business logic. \textbf{Compilability}: $F$ must be compilable by the Solidity compiler, i.e., $\text{Compilable}(C \cup F) = \text{True}$, adhering to language syntax, type rules, and correct integration with the contract context $C$. \textbf{Security}: $F$ must not contain known security vulnerability patterns, i.e., $\text{Secure}(F) = \text{True}$. This includes but is not limited to: (1) reentrancy protection through proper state updates before external calls, (2) access control enforcement via appropriate modifiers and permission checks, (3) integer overflow/underflow prevention using SafeMath or Solidity 0.8+ built-in protection, (4) proper validation of external inputs and contract calls, (5) avoidance of dangerous patterns such as unprotected \texttt{delegatecall}, timestamp dependencies, and unchecked low-level calls, and (6) adherence to established security patterns like checks-effects-interactions to prevent common attack vectors.

We propose a reasoning-enhanced approach that generates a function $\hat{F}$ and a reasoning chain $\hat{T}$. This ensures $\hat{F}$ meets all constraints, while $\hat{T}$ provides a transparent, auditable justification for the design choices, allowing developers to verify the security logic.

\subsection{Motivations}

\textbf{Summary.} Smart contract development demands rigorous security due to the immutable and high-stakes nature of deployed assets \cite{yu2024smart, yu2023money, yu2025smart, yu2023pscvfinder, yu2025sael}. However, existing Code LLMs often fail to address this, treating contract generation as a standard coding task without the necessary structured reasoning. This oversight leads to two critical limitations that motivate our work: (1) the absence of explicit, auditable security reasoning, leaving developers with unverifiable "black box" code, and (2) the resulting generation of critical vulnerabilities stemming from a shallow understanding of contract-specific principles like fund conservation.

\begin{figure}[htbp]
\centerline{\includegraphics[width=0.7\textwidth]{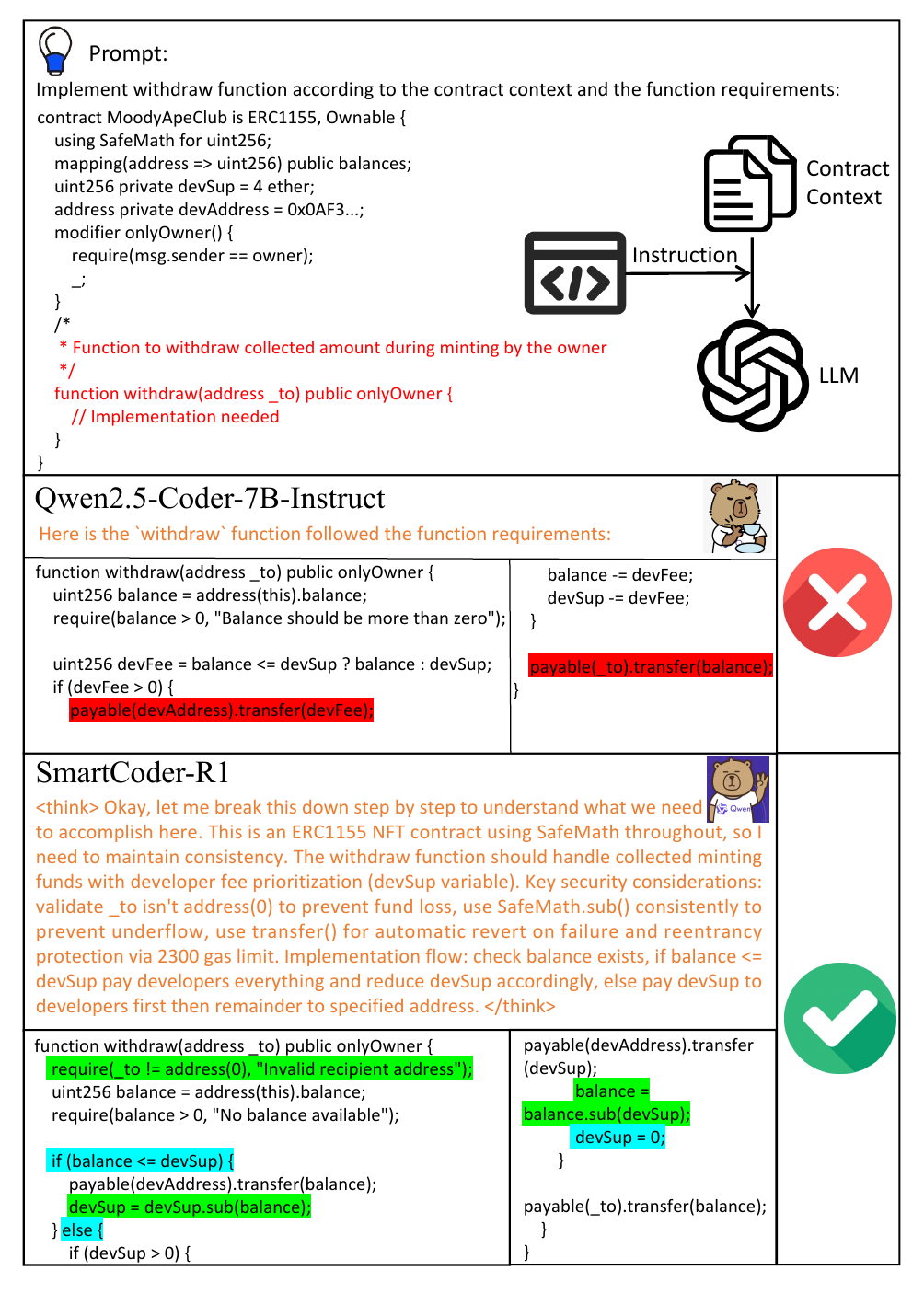}}
\caption{A motivating example to illustrate the limitations of non-reasoning Code LLMs in secure smart contract generation, demonstrating the advantages of reasoning-enhanced Code LLMs.}
\label{tab:motivation}
\end{figure}

\textbf{Motivation 1: Lack of Explicit Security Reasoning.} Most existing Code LLMs lack reasoning steps and simply treat smart contract generation as traditional code generation tasks. However, the immutable nature of smart contracts after deployment indicates that they require a higher level of security than traditional code generation tasks, as vulnerabilities can cause serious financial losses \cite{yu2023pscvfinder}. Therefore, smart contract generation tasks need to consider not only functional requirements but also security requirements. 

Security reasoning in smart contract development involves multi-layered thinking processes: threat modeling requires identifying potential attack vectors and malicious actors; state analysis demands understanding the security implications of contract state changes; interaction pattern analysis needs to consider secure interactions with other contracts and external calls; economic incentive analysis requires evaluating the security implications of tokenomics and incentive mechanisms. Existing code generation models such as the Qwen2.5-Coder series and CodeLLaMA series lack this structured reasoning capability and often generate code directly without demonstrating their security decision-making processes. This "black box" approach is perilous. When a model does not articulate its reasoning, developers cannot verify the security assumptions behind the code. As shown in Fig. \ref{tab:motivation}, Qwen2.5-Coder-7B-Instruct directly outputs code implementation without any reasoning process, making it impossible for developers to understand the security design principles and potential risks of the code. This "black box" generation process is particularly dangerous in high-risk blockchain environments, because smart contracts are usually immutable once deployed, and any security flaws can lead to permanent financial losses. In contrast, SmartCoder-R1 demonstrates its security considerations to developers through explicit reasoning processes: "need to use SafeMath to prevent underflow, use transfer() to automatically handle reentrancy attacks, implement developer fee priority logic to meet business requirements, ensure fund conservation principles." This transparent reasoning process enables programmers to understand, verify, and audit the security of generated code rather than blindly trusting model outputs.

\textbf{Motivation 2: Critical Security Vulnerabilities in Generated Code.} Taking the MoodyApeClub contract withdraw function implementation shown in Fig. \ref{tab:motivation} as an example, the code generated by Qwen2.5-Coder-7B-Instruct appears functionally correct but contains subtle yet critical business logic errors. Specifically, the code has a fund double payment vulnerability: the code first calculates the developer fee \texttt{uint256 devFee = balance <= devSup ? balance : devSup}, then pays the fee to the developer through \texttt{payable(devAddress).transfer(devFee)}, but the critical issue is that the code subsequently executes \texttt{payable(\_to).transfer(balance)}, transferring the original complete balance to the recipient without deducting the devFee already paid to the developer from the balance, resulting in total payments exceeding the funds actually held by the contract.

This error manifests security failures at three levels: fund integrity violation, where total transfer amounts exceed available balances and may cause transaction failures or fund over-issuance; business logic error, violating the expected business model of developer fee priority payment; state consistency breakdown, where contract state mismatches actual fund allocation. In the MoodyApeClub NFT project revenue distribution scenario, this error would cause developers to receive fees while artists and promoters can still extract full shares, thereby depleting contract funds. It is particularly noteworthy that such errors often pass basic syntax checks and simple functional tests because the code structure appears reasonable, but in complex business scenarios can lead to serious financial losses. Qwen2.5-Coder-7B-Instruct failed to understand contextual requirements and priority relationships between different stakeholders in the smart contract ecosystem.

In contrast, SmartCoder-R1 identified this potential problem through explicit security reasoning processes. In its reasoning process, the model explicitly mentioned the need to "implement developer fee priority logic to meet business requirements, ensure fund conservation principles." Based on this understanding, SmartCoder-R1 generated the correct implementation: handling different situations through \texttt{if (balance <= devSup)} conditional branches, using \texttt{devSup = devSup.sub(balance)} to update remaining developer fees in the first branch, using \texttt{balance = balance.sub(devSup)} to correctly deduct paid developer fees from the balance in the second branch, and finally executing \texttt{payable(\_to).transfer(balance)} to transfer the updated balance. This implementation ensures fund conservation principles and avoids double payment vulnerabilities.

This case clearly demonstrates the limitations of current code generation models: lack of deep understanding of constraints specific to smart contracts (such as fund conservation, state consistency, multi-party interest balance). SmartCoder-R1 not only avoids fund allocation errors but also explicitly articulates the security considerations of its design decisions. This comparison clearly shows why we need a smart contract generation model capable of explicit security reasoning, and why existing direct code generation approaches are insufficient in this security-critical domain. This conclusion is further substantiated by our empirical data in Table \ref{tab:rq1_main}. A direct comparison reveals a clear performance gap: the reasoning-enhanced DeepSeek-R1 achieves a FullRate of 34.66\%, significantly outperforming its standard counterpart, DeepSeek-V3 (30.16\%). This pattern holds for the Qwen family, where the reasoning-based QwQ-32B (23.94\%) surpasses the standard Qwen2.5-32B-Instruct (20.50\%). These quantitative results provide strong empirical evidence that models with explicit reasoning capabilities produce more functional and secure code.

\section{Approach}

\begin{figure*}[htbp]
  \centering
  \includegraphics[width=1\textwidth]{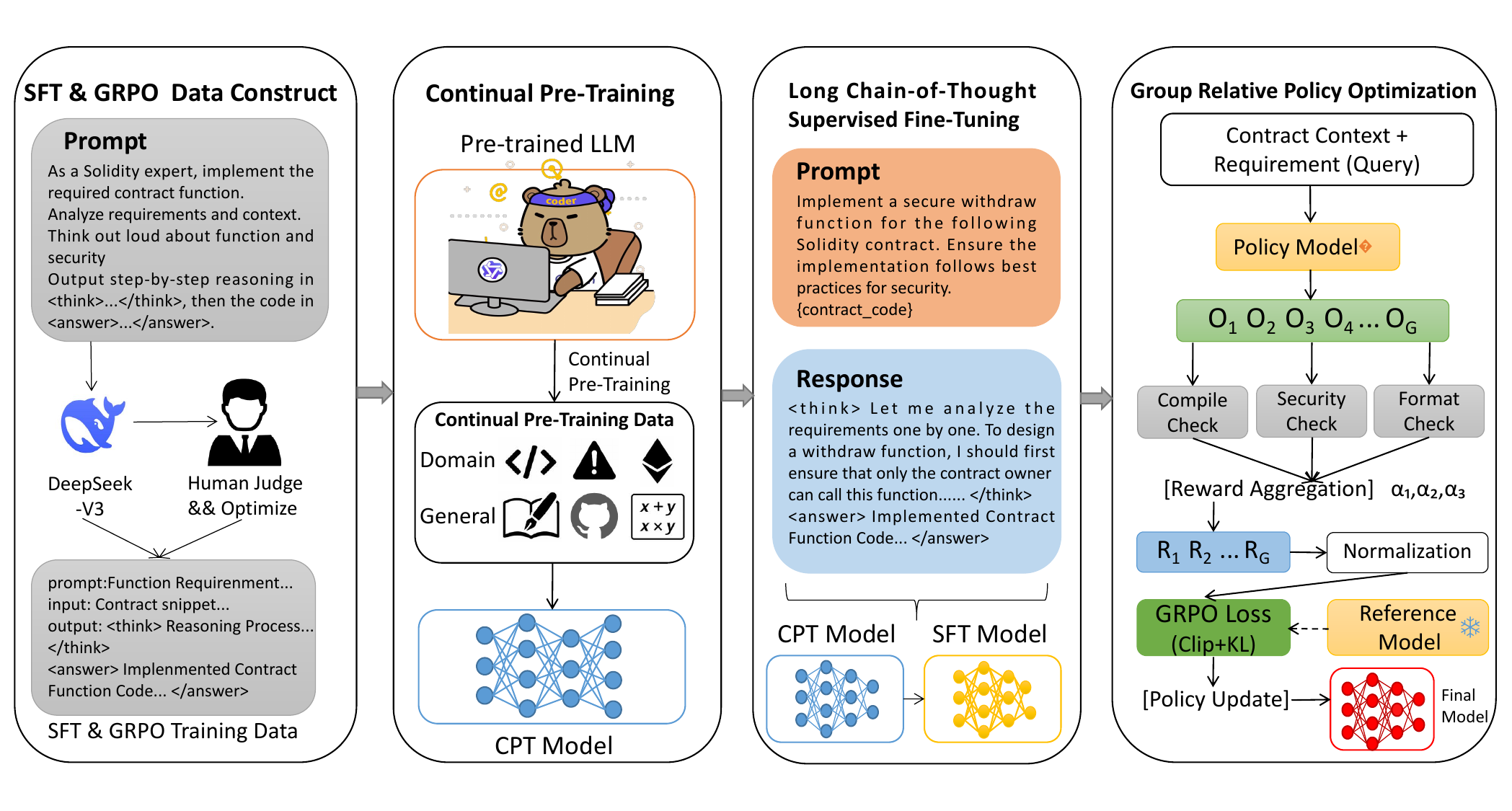}
  \caption{Overview of our SmartCoder-R1 pipeline.}
  \label{overview}
\end{figure*}

\textbf{Summary.}
The \textit{Approach} section details our end-to-end methodology for Dataset Construction, CPT, L-COT SFT, and Reinforcement Learning with S-GRPO. The Qwen2.5-Coder-7B serves as our base model, selected for its superior coding capabilities \cite{hui2024qwen2} and the Qwen family's superior adaptability to reinforcement learning (RL) compared to counterparts like Llama \cite{wang2025octothinker}. As shown in Fig. \ref{overview}, we begin by building a high-quality dataset with explicit vulnerability annotations and multi-step reasoning traces, followed by continual pre-training on normalized Solidity code to instill syntactic and structural knowledge. Next, we employ Long Chain-of-Thought Supervised Fine-Tuning to teach the model explicit reasoning about security requirements. Finally, we apply S-GRPO reinforcement learning to further align the model with rigorous security, compliance, and functionality standards, leveraging automated reward signals and KL-regularized policy updates.

\subsection{Dataset Construction}

We construct the SFT and S-GRPO datasets through a multi-step process that combines deterministic programmatic extraction, LLM-based data generation, and manual expert validation. The raw contract data is sourced from the Ethereum mainnet, filtered to include only contracts with verified source code and explicit function-level documentation. For each contract, we parse the class-level and function-level code, extracting the function signature, documentation string, and the entire Solidity context required for correct compilation. Where function documentation is in Chinese, we translate it into English using DeepSeek-V3 with a translation prompt, and fallback to a pattern-based description generator if translation fails. All documentation is cleaned to remove escape characters, annotation tags, and non-informative boilerplate.

For each function, we generate an instruction that specifies the target function name and the associated requirements. We then construct an LLM prompt that includes the full contract context, the function name, the cleaned and translated documentation, and a reference implementation if available. We use DeepSeek-V3 with a conversational, chain-of-thought prompt to generate a sample that includes a detailed, stepwise reasoning process in \texttt{<think>...</think>} tags and a standalone, syntactically correct Solidity implementation in \texttt{<answer>...</answer>} tags. The LLM is instructed to consider twelve specific vulnerability classes, including reentrancy, delegatecall, unchecked call, price oracle manipulation, unprotected suicide, integer overflow/underflow for Solidity versions before 0.8.0, timestamp dependency, access control, and others.

Each LLM-generated sample is post-processed to extract the reasoning chain (via regex for \texttt{<think>} tags), the implementation code (from \texttt{<answer>} tags), and to identify which security-sensitive patterns are actually present in the code (using fixed regular expressions for Solidity constructs and keywords). All samples are then reviewed by human experts with experience in Solidity security auditing. Reviewers verify that the generated code can be compiled by the Solidity compiler (\texttt{solc}), that the code addresses all stated requirements, and that it does not introduce any of a pre-defined set of security vulnerabilities. The expert review process includes both manual reading and static analysis through professional-grade security scanning tools. Samples that do not compile, fail security checks, or exhibit incomplete reasoning are either corrected or removed. All finalized samples are used as instruction-response pairs for SFT, and as multi-candidate, reward-annotated tuples for S-GRPO.

\subsection{Continual Pre-training}

During continual pre-training, we use only pure Solidity code, excluding comments and documentation, to prevent the model from memorizing non-functional text and to focus the learning on syntactic and structural patterns. Each contract is decomposed into a normalized format with explicit separation of contract-level declarations, state variables, modifiers, events, and function implementations. We segment each contract into overlapping code windows of up to 2048 tokens. Each window is used as a training instance for next-token prediction.

The training objective is the standard left-to-right language modeling loss:
\begin{align}
\mathcal{L}_{\text{CPT}}
= -\frac{1}{N} \sum_{i=1}^{N} \sum_{j=1}^{T_i} \log P(x_{i,j} \mid x_{i,<j})
\label{eq:cpt}
\end{align}
where $N$ is the total number of code windows in the pre-training corpus, $T_i$ is the length of the $i$-th window, and $x_{i,j}$ is the $j$-th token in window $i$. We train the model for two epochs using the AdamW optimizer with a learning rate of $1 \times 10^{-5}$, batch size of 64, and a sequence length cutoff of 2048 tokens. Model checkpoints are saved every 500 steps. During pre-training, the model is exposed to contract patterns including inheritance, state variable initialization, error handling via \texttt{require} and \texttt{revert}, and common access control idioms. No reasoning or documentation information is present in this phase.

\subsection{Long Chain-of-Thought Supervised Fine-Tuning}

The purpose of the Long Chain-of-Thought (CoT) Supervised Fine-Tuning stage is to explicitly endow the model with the capacity to perform multi-step reasoning required for secure smart contract code generation. Each training sample consists of an instruction, contract context, a detailed stepwise reasoning chain enclosed in \texttt{<think>...</think>}, and a standalone Solidity implementation enclosed in \texttt{<answer>...</answer>}. The output sequence is strictly required to begin with a multi-sentence, logically ordered reasoning trace, followed by a complete function implementation. The supervised fine-tuning loss is:
\begin{align}
\mathcal{L}_{\text{SFT}}
= -\frac{1}{M} \sum_{k=1}^{M} \sum_{l=1}^{L_k} \log P(y_{k,l} \mid I_k, X_k, y_{k,<l})
\label{eq:sft}
\end{align}
where $M$ is the number of samples, $I_k$ is the instruction, $X_k$ is the input, $y_{k,l}$ is the $l$-th output token. The model is trained for three epochs, learning rate $1 \times 10^{-5}$, batch size 8, and maximum sequence length 8192. All outputs are required to compile with \texttt{solc} and to address at least one specified vulnerability class. This stage imparts the model with the initial ability to decompose business requirements into explicit security logic steps and to realize these as functional, secure, and compilable code. After SFT, the model can generate code that is not only correct and complete, but also accompanied by an auditable reasoning trace, providing the foundation for downstream RL alignment.

\subsection{Security-Aware Group Relative Policy Optimization}

The S-GRPO stage acts as a targeted refinement process, aligning the model with the complex, multi-faceted objectives of real-world smart contract development that are difficult to capture with supervised learning alone. For each input, which consists of a contract's existing code (\texttt{context}) and a new functional requirement, the model's task is to generate a response that contains both: (1) a \textbf{Long Chain-of-Thought} (\texttt{<think>}) reasoning process that explicitly outlines the security considerations and implementation plan, and (2) the final, compilable \textbf{Solidity function} (\texttt{<answer>}).

During training, we sample a group of $G$ candidate outputs from the current policy network for each prompt. Each candidate is then subjected to a rigorous, automated evaluation to calculate its reward:

\begin{itemize}
    \item \textbf{Compilation Check ($R_{\text{compile}}$):} This is the most foundational prerequisite for any smart contract. The generated Solidity code is compiled using \texttt{solc}. A successful compilation ($R_{\text{compile}}=1$) signifies that the code is syntactically valid and can be deployed on the blockchain. A failure ($R_{\text{compile}}=0$) indicates a fundamental flaw, rendering the output useless.

    \item \textbf{Security Rule Check ($R_{\text{security}}$):} This is the most critical component, directly targeting the core challenge of secure contract generation. We use a combination of regular expressions and static analysis scripts to scan both the reasoning and the code for known vulnerability patterns. It is crucial to note that these checks are intentionally designed to be conservative, operating on a "better safe than sorry" principle. This means the system prioritizes flagging any potentially suspicious code patterns, even if it results in occasional false positives (flagging safe code as risky), rather than risking a single false negative (missing a true vulnerability). For example, when implementing a \texttt{withdraw} function, this check verifies if the model correctly applies the \textbf{Checks-Effects-Interactions (CEI)} pattern to mitigate reentrancy attacks. It also checks for missing access control modifiers (e.g., ensuring a critical function is marked \texttt{onlyOwner}), the unsafe use of \texttt{delegatecall}, and potential integer overflows/underflows in contracts targeting versions prior to Solidity 0.8.0, where such checks are necessary. A secure output receives $R_{\text{security}}=1$; otherwise, it receives $R_{\text{security}}=0$.

    \item \textbf{Format Compliance ($R_{\text{format}}$):} This reward ensures the model maintains its explainability, a cornerstone of our approach. The output must contain a well-structured \texttt{<think>} block with substantive reasoning (at least three steps) followed by a distinct \texttt{<answer>} block. This structure provides a crucial \textbf{cognitive scaffold} for developers, allowing them to audit the model's security logic before trusting its code. A compliant format yields $R_{\text{format}}=1$.
\end{itemize}

The total reward $R$ for each candidate is a weighted sum of these three binary scores, reflecting their relative importance:
\begin{align}
R = \alpha_1 R_{\text{compile}} + \alpha_2 R_{\text{security}} + \alpha_3 R_{\text{format}}
\end{align}
We set the weights to $(\alpha_1, \alpha_2, \alpha_3) = (0.3, 0.5, 0.2)$, prioritizing security ($R_{\text{security}}$) above all else, as a single vulnerability can compromise an entire contract.

The policy is updated using the S-GRPO objective, which combines the clipped policy ratio surrogate loss with a KL penalty. The overall S-GRPO loss is:
\begin{align*}
J_{\mathrm{S-GRPO}}(\theta) =\; & \mathbb{E}_{q, \{o_i\}_{i=1}^G} \Bigg[
    \frac{1}{G} \sum_{i=1}^G \frac{1}{|o_i|} \sum_{t=1}^{|o_i|} \min\left(
        \frac{\pi_\theta(o_{i,t}|q, o_{i,<t})}{\pi_{\theta_{\text{old}}}(o_{i,t}|q, o_{i,<t})} \hat{A}_{i,t}, \right. \\
    &\qquad \left. \mathrm{clip}\left(\frac{\pi_\theta(o_{i,t}|q, o_{i,<t})}{\pi_{\theta_{\text{old}}}(o_{i,t}|q, o_{i,<t})}, 1-\epsilon, 1+\epsilon\right) \hat{A}_{i,t}
    \right)
\Bigg]
- \beta D_{\mathrm{KL}}[\pi_\theta\|\pi_{\text{ref}}] \tag{4}
\end{align*}
where $\hat{A}_{i,t}$ is the normalized advantage, and $\beta$ is the KL penalty coefficient. The KL divergence $D_{\mathrm{KL}}[\pi_\theta\|\pi_{\text{ref}}]$ is estimated for every token in the output sequence using an unbiased estimator.

Concretely, during each RL step, the model receives as input the full contract context and functional requirement, and must generate a reasoning chain that explicitly discusses security best practices and then output a function that is both functionally correct and secure. The reward system ensures that the model only receives a high score if the code passes full compilation, contains no statically detectable vulnerabilities, and the reasoning is explicit and well-structured. RL training is performed for 5 epochs, with all outputs and reward components logged.

In essence, the S-GRPO method constructs an efficient, automated "trial-and-error" learning environment for the model, with the core idea of teaching it to write secure smart contracts through group-based comparison and immediate feedback. In practice, when faced with a specific programming task (e.g., implementing a re-entrancy-proof \texttt{withdraw} function), the model first generates a diverse group of potential solutions. Some of these attempts may be perfect, others might contain security vulnerabilities, and some may not even compile. The system then acts as an automated "code reviewer," rigorously evaluating each attempt: solutions that compile successfully, adhere to secure design patterns (like the Checks-Effects-Interactions principle), and provide clear reasoning are given a high reward (a high advantage). Conversely, those with vulnerabilities or errors receive a low reward or penalty (a low advantage). This reward signal directly guides the model's parameter updates, reinforcing the behaviors that lead to high-quality outputs while discouraging those that result in problems. Through this continuous, automated iteration loop, S-GRPO compels the model to not only learn "how to write code," but more importantly, to internalize the core development principles of "why it must be written this way to be secure."

\section{Experiments}

\subsection{Research Questions}
To evaluate our SmartCoder-R1, we conduct experiments to answer the following research questions:

$\bullet$ $\textbf{RQ1}$: How does our SmartCoder-R1 perform in generating secure and functional smart contracts compared to state-of-the-art code generation models?

$\bullet$ $\textbf{RQ2}$: How do individual components contribute to the overall performance, and how sensitive is performance to key parameter tuning?

$\bullet$ $\textbf{RQ3}$: How effective are the reasoning chains generated by SmartCoder-R1 in terms of their Functionality, Security, and Clarity?

$\bullet$ $\textbf{RQ4}$: How does SmartCoder-R1’s reasoning and security awareness compare to state-of-the-art methods through representative case studies?

$\bullet$ $\textbf{RQ5}$: What are the primary vulnerability types in generated insecure contracts, and what factors contribute to these security failures?

\subsection{Dataset}

\textbf{Continual Pre-training (CPT):} Following Yu et al.\cite{yu2025smart}, we construct our CPT dataset from 186,397 unique Ethereum smart contracts provided by Storhaug et al.\cite{storhaug2023efficient} (501.62M tokens). To enhance diversity, we augment this with 100,000 instances of general code, mathematics, and texts in English and Chinese (118.94M tokens), creating a corpus of 286,397 instances totaling 620.56M tokens.

\textbf{Long Chain-of-Thought SFT (L-COT SFT):} Our SFT dataset features 7,998 meticulously annotated samples (16.44M tokens, avg. 2,056 per sample) from 1,508 unique contracts from Storhaug et al.~\cite{storhaug2023efficient}. It covers diverse patterns, including ERC20/721, Ownable, crowdsales, pausable/proxy patterns, AccessControl, multisig wallets, Governor/Timelock modules, and DeFi primitives. Crucially, each instance contains the full contract context, functional requirements, the target implementation, and an explicit Chain-of-Thought reasoning trace constructed by us. The function distribution is realistic (constructors 4.5\%, regular 55.3\%, view 35.2\%, payable 5.1\%). Correctness was ensured through both manual verification and automated analysis.

\textbf{Security-Aware Group Relative Policy Optimization (S-GRPO):} The S-GRPO dataset is built from the same sources as the SFT stage but is strictly separated. It contains 1,691 samples (2.74M tokens, avg. 1,621 per sample), with 1,200 contracts for training and 491 for evaluation. Notably, we constructed new reasoning chains tailored for this RL stage. Each sample provides a contract context, requirement, reasoning chain, and labeled implementation to facilitate policy optimization with step-by-step feedback.

\textbf{Evaluation:} Our evaluation set comprises 756 unique functions from 289 real-world contracts, sourced from Storhaug et al.~\cite{storhaug2023efficient} and EtherScan, spanning 53 Solidity versions. It includes 335 distinct implementations, with complexity ranging from 44-character utilities to 3,149-character business logic. The dataset robustly covers multi-layered structures, from libraries (SafeMath, EnumerableSet, SafeCast) to critical patterns (Ownable, Proxy, Roles), and features 88.10\% high-quality NatSpec documentation.

\noindent\textbf{Data Leakage Prevention:} To ensure rigorous evaluation, we enforced strict non-overlapping splits between CPT, SFT, S-GRPO, and Evaluation datasets. At the function level, we applied token-based Jaccard similarity deduplication (threshold 0.9) to remove near-duplicates. The evaluation set was completely isolated from all training and fine-tuning data, guaranteeing zero overlap and preventing data leakage.

\subsection{Baselines}


We evaluate our model against state-of-the-art baselines from three categories: \textbf{General/Code LLMs}: These are LLMs with strong, but not specialized, code generation capabilities. This category includes: the CodeLlama-Instruct series (7B, 13B, 34B)\cite{roziere2023code}, the DeepSeek-Coder-Instruct series (6.7B, 33B)\cite{guo2024deepseek}, the Qwen-Coder series (Qwen2.5-3B, 14B, 32B-Instruct~\cite{hui2024qwen2}; Qwen3-30B-A3B-Instruct~\cite{yang2025qwen3}), Qwen2.5-32B-Instruct~\cite{team2024qwen2}, DeepSeek-V3~\cite{liu2024deepseek}, the Llama-3 series (3.1-8B, 3.2-1B), and the commercial models GPT-4.1~\cite{openai_gpt41_2025}. \textbf{Domain-Specific Smart Contract LLMs}: These are LLMs specifically designed for smart contract generation. We include Storhaug et al.\cite{storhaug2023efficient}. \textbf{Reasoning LLMs}: These LLMs incorporate explicit multi-step reasoning. We compare against DeepSeek-R1~\cite{guo2025deepseek} and QwQ-32B~\cite{qwen_qwq_2025}, which serve as strong baselines for reasoning-enhanced generation.

\subsection{Metrics}

Following established evaluation frameworks \cite{wang2025codebc}, we use five primary metrics. We measure \textbf{ComPass(\%)}, the proportion of successfully compiled code. As a conditional metric, \textbf{VulRate(\%)} is the percentage of compilable code containing known vulnerabilities (lower is better). We then use three "yield" metrics: \textbf{SafeAval(\%)} for code that is both compilable and secure, \textbf{FuncRate(\%)} for code that is compilable and functionally correct, and \textbf{FullRate(\%)}, the proportion meeting all criteria. To rigorously evaluate \textbf{VulRate} and \textbf{SafeAval}, we combined automated scanning with expert auditing. First, all successfully compiled code was scanned with the \textbf{Slither} tool, which categorizes potential vulnerabilities by severity (e.g., High, Medium, Low). Due to the high cost of expert audits, we selected a sample of \textbf{100} code snippets for manual review. This selection prioritized snippets with High-severity warnings, followed by those with Medium-severity. The final \textbf{VulRate} is a calibrated metric. To evaluate functional correctness (\texttt{FuncRate}), we employ an execution-based validation framework. For each function in our evaluation set, we first developed a comprehensive unit test suite based on its ground-truth oracle. This suite is designed to verify all critical aspects of the function's behavior. The generated code from a model is then programmatically integrated into the contract, and the exact same test suite is executed against it. FuncRate is then calculated as the percentage of generated functions that successfully pass their entire corresponding test suite.

To evaluate the quality of the generated reasoning chains (RQ3), we manually assessed them on a 4-point Likert scale \cite{joshi2015likert} across three dimensions: Functionality, which measures if the reasoning correctly understands and plans for all required features; Security, which assesses whether all relevant security risks are identified and addressed in the plan; and Clarity, which evaluates the logical flow and readability of the explanation. To ensure the objectivity and reliability of our manual evaluation, the assessment was conducted by six senior experts, each with over five years of experience in Solidity development and security auditing. A rigorous assurance mechanism was implemented to minimize subjectivity. Initially, all six experts calibrated their understanding by independently rating a pilot set of 30 samples. Inter-rater reliability for this pilot set was calculated using Fleiss' Kappa, yielding a coefficient of 0.85, which indicates substantial agreement. For the formal assessment, each sample was assigned to a team of two experts for independent scoring. In cases where the two experts' scores had a significant discrepancy (a difference of 2 or more points Likert scale), a third expert arbitrated the result to reach a final, consensus-based rating.

We deliberately do not use traditional code similarity metrics (e.g., BLEU \cite{papineni2002bleu}, CodeBLEU \cite{ren2020codebleu}). These metrics measure surface-level resemblance to a reference, which is an insufficient proxy for quality in smart contracts, where security and functional correctness are paramount and multiple valid implementations can exist. We avoid sampling-based metrics like pass@k, as they tolerate the generation of insecure candidates, an unacceptable risk in a domain that demands deterministic reliability over probabilistic correctness in smart contracts.

\subsection{Implementation Details}

We perform CPT and L-COT SFT using LlamaFactory~\cite{zheng2024llamafactory} and DeepSpeed~\cite{rasley2020deepspeed}, both with \texttt{fp16} enabled. Loss is calculated using cross-entropy, and parameters are optimized with AdamW~\cite{adamw} (\(\beta = (0.9, 0.99)\), \(\epsilon = 1\mathrm{e}{-8}\)). For all our models, we employ full parameter tuning. \textbf{During CPT}, the batch size is \(64\) per device, gradient accumulation steps are \(16\), epochs are \(2\), learning rate is \(1\mathrm{e}{-5}\) with cosine decay, warmup steps are \(0\), cutoff length is \(2048\), and save steps are \(500\). \textbf{During L-COT SFT}, the batch size is \(8\) per device, gradient accumulation steps are \(8\), epochs are \(3\), learning rate is \(1\mathrm{e}{-5}\) with cosine decay, warmup steps are \(0\), cutoff length is \(8192\), and save steps are \(50\). \textbf{For S-GRPO}, we build our implementation upon the Logic-RL~\cite{xie2025logic} and VeRL~\cite{sheng2025hybridflow}. We initialize from the L-COT SFT checkpoint and use a batch size of \(8\) per device, a maximum prompt length of \(24{,}576\), maximum response length of \(2{,}048\), and a learning rate of \(3 \times 10^{-7}\). KL loss with low variance is employed as a regularization term (\(\text{coef}=0.001\)). Both the mini-batch and micro-batch sizes are set to \(8\). Gradient checkpointing and full parameter, gradient, and optimizer offloading are enabled to maximize GPU memory utilization. The number of parallel rollouts is set to \(8\) (\texttt{rollout.n=8}). Rollout and reference log probability micro-batch sizes are set to \(160\). Training is conducted on a single node with \(8\) NVIDIA H800 GPUs (each with \(80\)GB memory), for \(5\) epochs, saving checkpoints every \(20\) steps. \textbf{For Evaluation}, we use greedy decoding (\texttt{do\_sample=false}) to ensure stable and reproducible results. This approach eliminates the randomness introduced by sampling-based decoding strategies, thereby allowing for fair comparison across different models and experimental runs. All evaluation experiments are conducted on the same hardware to ensure consistency.

\subsection{Experimental Results}
In this section we present experimental results and analysis to answer the research question.

1) RQ1: As shown in Table \ref{tab:rq1_main}, SmartCoder-R1 achieves the highest overall performance across all key metrics when evaluated against a comprehensive range of baselines.

Specifically, SmartCoder-R1 achieves a ComPass (compilability) of 87.70\%, improving over the best general/code LLM (GPT-4.1 at 84.13\%) by 4.24\% relatively and over the best domain-specific model (Storhaug et al. at 82.67\%) by 6.08\% relatively. On the critical VulRate metric (lower is better), SmartCoder-R1 attains a remarkably low vulnerability rate of 8.60\%, which is 49.05\% lower than CodeBC (16.88\%), 44.80\% lower than Qwen2.5-Coder-32B-Instruct (15.58\%), and 72.09\% lower than GPT-4.1 (30.82\%). For the SafeAval (compilable and secure) metric, SmartCoder-R1 achieves 80.16\%, representing a 17.20\% relative improvement over CodeBC (68.39\%) and a 26.52\% relative improvement over Storhaug et al. (63.36\%). Furthermore, SmartCoder-R1 demonstrates notable improvements in FuncRate (functional correctness, 53.84\%), slightly outperforming the strongest baselines (e.g., +1.76\% relative over GPT-4.1 at 52.91\%). Especially, SmartCoder-R1 reaches a FullRate (functional correctness, security, and compilability simultaneously) of 50.53\%, reflecting a 45.79\% relative improvement over the next best result (DeepSeek-R1 at 34.66\%). The seemingly low VulRate of some smaller models is an artifact of their frequent failure to produce compilable code (e.g., incomplete snippets or plain text), which are excluded from analysis.

\begin{table*}[ht]
    \centering
    \caption{RQ1: Comparison of Secure and Functional Smart Contract Generation. Performance (\%) of SmartCoder-R1 and state-of-the-art code generation baselines on secure smart contract generation benchmarks. $\uparrow$ indicates higher is better, $\downarrow$ indicates lower is better. \textbf{Bold} highlights the best results in each column.}
    \label{tab:rq1_main}
    \resizebox{\textwidth}{!}{
    \begin{tabular}{l|c|c|c|c|c}
        \toprule
        \textbf{Model} & \textbf{ComPass} $\uparrow$ & \textbf{VulRate} $\downarrow$ & \textbf{SafeAval} $\uparrow$ & \textbf{FuncRate} $\uparrow$ & \textbf{FullRate} $\uparrow$ \\
        \midrule
        \multicolumn{6}{c}{\textbf{\textit{General / Code LLMs}}} \\
        \midrule
        CodeLlama-7b-Instruct      & 29.63 & 14.29 & 25.40 & 13.62 & 12.57 \\
        CodeLlama-13b-Instruct     & 54.63 & 23.24 & 41.93 & 25.13 & 21.83 \\
        CodeLlama-34b-Instruct     & 48.94 & 25.14 & 36.64 & 23.15 & 20.63 \\
        Llama-3.2-1B-Instruct      & 7.01 & 16.98 & 5.82 & 3.17 & 1.98 \\
        Llama-3.1-8B-Instruct      & 22.09 & 11.98 & 19.44 & 6.35 & 5.29 \\
        DeepSeek-Coder-6.7B-Instruct & 15.08 & 21.05 & 11.90 & 4.89 & 4.37 \\
        DeepSeek-Coder-33B-Instruct  & 18.52 & 25.71 & 13.76 & 12.30 & 10.58 \\
        Qwen2.5-Coder-3B-Instruct & 12.57 & 22.11 & 9.79 & 7.01 & 5.29 \\
        Qwen2.5-Coder-14B-Instruct & 51.19 & 10.85 & 45.63 & 24.34 & 20.90 \\
        Qwen2.5-Coder-32B-Instruct & 62.83 & 15.58 & 53.04 & 38.23 & 29.50 \\
        Qwen2.5-32B-Instruct & 48.02 & 22.31 & 37.30 & 27.12 & 20.50 \\
        Qwen3-Coder-30B-A3B-Instruct   & 78.70 & 37.14 & 49.47 & 41.40 & 19.97 \\
        DeepSeek-V3                & 69.05 & 33.91 & 45.63 & 48.28 & 30.16 \\
        GPT-4.1                    & 84.13 & 30.82 & 58.20 & 52.91 & 30.29 \\
        \midrule
        \multicolumn{6}{c}{\textbf{\textit{Domain-Specific Smart Contract LLMs}}} \\
        \midrule
        Storhaug et al. \cite{storhaug2023efficient}     & 82.67 & 23.36 & 63.36 & 34.79 & 23.54 \\
        \midrule
        \multicolumn{6}{c}{\textbf{\textit{Reasoning LLMs}}} \\
        \midrule
        DeepSeek-R1                & 84.39 & 32.29 & 57.14 & 50.66 & 34.66 \\
        QwQ-32B                    & 67.46 & 44.12 & 37.70 & 40.48 & 23.94 \\
        \midrule
        \textbf{SmartCoder-R1 (Ours)} & \textbf{87.70} & \textbf{8.60} & \textbf{80.16} & \textbf{53.84} & \textbf{50.53} \\
        \bottomrule
    \end{tabular}
    }
\end{table*}

\vspace{5pt}
\noindent\begin{tikzpicture}
  \node[draw=black, thick, fill=gray!20, rounded corners, inner sep=10pt, text width=0.92\linewidth] {
    \textbf{Answer to RQ1:} SmartCoder-R1 achieves state-of-the-art performance in generating secure smart contracts that meet functional requirements, outperforming the strongest baseline across all key metrics (ComPass, VulRate, SafeAval, FuncRate, FullRate).
  };
\end{tikzpicture}
\vspace{5pt}

2) RQ2: Our ablation study in Table \ref{tab:ablation} systematically dissects the contribution of each core component: Continual Pre-training (CPT), Long Chain-of-Thought Supervised Fine-Tuning (L-COT SFT), and Security-Aware Group Relative Policy Optimization (S-GRPO) to the overall performance.

\begin{table}[ht]
    \centering
    \caption{Ablation Study: Impact of Different Training Components. Performance (\%) of SmartCoder-R1 variants with different training configurations. w/o: without.}
    \label{tab:ablation}
    \begin{tabular}{l|c|c|c|c|c}
        \toprule
        \textbf{Model Variant} & \textbf{ComPass} $\uparrow$ & \textbf{VulRate} $\downarrow$ & \textbf{SafeAval} $\uparrow$ & \textbf{FuncRate} $\uparrow$ & \textbf{FullRate} $\uparrow$ \\
        \midrule
        w/o CPT \& S-GRPO & 80.82 & 19.80 & 64.81 & 31.22 & 20.37 \\
        w/o CPT \& L-COT SFT & 24.60 & 9.68 & 22.22 & 15.08 & 14.55 \\
        w/o CPT & 85.05 & 14.15 & 73.02 & 37.17 & 31.75 \\
        w/o S-GRPO & 84.26 & 18.68 & 68.52 & 32.67 & 24.21 \\
        w/o L-COT SFT & 67.86 & 27.10 & 49.47 & 46.96 & 32.67 \\
        Full Model & \textbf{87.70} & \textbf{8.60} & \textbf{80.16} & \textbf{53.84} & \textbf{50.53} \\
        \bottomrule
    \end{tabular}
\end{table}

\begin{table}[htbp]
\centering
\caption{Parameter Sensitivity Analysis Results.}
\label{tab:sensitivity_analysis}
\resizebox{\textwidth}{!}{%
\begin{tabular}{l|c|ccccc}
\toprule
\textbf{Configuration} & \textbf{Weights} & \textbf{ComPass} & \textbf{VulRate} & \textbf{SafeAval} & \textbf{FuncRate} & \textbf{FullRate} \\
 & \textbf{($\alpha_1, \alpha_2, \alpha_3$)} & \textbf{(\%) $\uparrow$} & \textbf{(\%) $\downarrow$} & \textbf{(\%) $\uparrow$} & \textbf{(\%) $\uparrow$} & \textbf{(\%) $\uparrow$} \\
\midrule
Security+ & (0.2, 0.6, 0.2) & 86.51 & 13.91 & 74.41 & 47.09 & 39.81 \\
Security++ & (0.1, 0.7, 0.2) & 85.71 & 15.28 & 72.62 & 45.90 & 36.64 \\
Compile+ & (0.4, 0.4, 0.2) & 85.85 & 16.02 & 72.09 & 49.21 & 41.53 \\
Compile++ & (0.5, 0.3, 0.2) & 85.32 & 15.66 & 71.96 & 54.10 & 45.77 \\
Compile+++ & (0.6, 0.2, 0.2) & 84.39 & 16.77 & 70.24 & 50.40 & 40.48 \\
Compile++++ & (0.7, 0.1, 0.2) & 85.71 & 19.14 & 69.31 & 49.21 & 39.55 \\
Ours & (0.3, 0.5, 0.2) & \textbf{87.70} & \textbf{8.60} & \textbf{80.16} & 53.84 & \textbf{50.53} \\
\bottomrule
\end{tabular}%
}
\end{table}

The analysis begins with the baseline model trained only with L-COT SFT (\texttt{w/o CPT \& S-GRPO}). While it achieves a respectable \texttt{ComPass} of 80.82\%, it is insufficient for security, with a high \texttt{VulRate} of 19.80\% and a low \texttt{FullRate} of 20.37\%. This highlights that teaching a reasoning-and-code format alone does not confer the deep understanding needed to avoid security pitfalls. When CPT is introduced (\texttt{w/o S-GRPO}), metrics improve (\texttt{ComPass} to 84.26\%, \texttt{SafeAval} to 68.52\%) due to a better understanding of Solidity. However, the \texttt{VulRate} remains high at 18.68\%, confirming this combination cannot fully solve the security challenge. The most significant leap in security comes from the S-GRPO stage. Comparing the full model to its \texttt{w/o S-GRPO} variant, the \texttt{VulRate} plummets from 18.68\% to an industry-leading 8.60\% (a 53.96\% relative reduction), boosting \texttt{SafeAval} from 68.52\% to 80.16\% and \texttt{FullRate} from 24.21\% to 50.53\%. This underscores S-GRPO's power to directly optimize for complex objectives like security and compilability.

An intriguing counterpoint is the \texttt{w/o L-COT SFT} variant (CPT + S-GRPO), which achieves a surprisingly high \texttt{FuncRate} (46.96\%), nearly matching the full model. This suggests that with a strong foundation from CPT, S-GRPO's powerful optimization can directly steer the model towards functional correctness. However, this approach comes at a steep cost: its \texttt{VulRate} is the highest among the primary variants at 27.10\%. This starkly illustrates the indispensable role of L-COT SFT. L-COT SFT does not just provide training data; it teaches the model a crucial, structured reasoning process (\texttt{<think>...</think>}). This reasoning framework acts as a cognitive scaffold, guiding the model to explicitly consider security constraints \textit{before} generating code, making the subsequent S-GRPO refinement far more effective and targeted. Without this scaffold, the model, while capable of finding functionally correct solutions, frequently fails to navigate complex security requirements, leading to insecure implementations. 

To further investigate the impact of S-GRPO, we conducted a parameter sensitivity analysis on the reward weights, with results presented in Table~\ref{tab:sensitivity_analysis}. This analysis reveals the delicate balance required between compilation ($\alpha_1$), security ($\alpha_2$), and format ($\alpha_3$). Configurations that over-prioritize security (e.g., `Security++' with $\alpha_2=0.7$) paradoxically result in a higher vulnerability rate (15.28\%) compared to our balanced approach (8.60\%). Conversely, increasing the weight on compilation (e.g., `Compile++' with $\alpha_1=0.5$) boosts functional correctness to 54.10\% but at the expense of security (VulRate of 15.66\%). Our chosen configuration ($\alpha_1=0.3, \alpha_2=0.5, \alpha_3=0.2$) strikes an optimal trade-off by effectively harmonizing all three objectives, achieving the highest FullRate (50.53\%).

\vspace{5pt} 
\noindent\begin{tikzpicture}
  \node[draw=black, thick, fill=gray!20, rounded corners, inner sep=10pt, text width=0.92\linewidth] {
    \textbf{Answer to RQ2:} The ablation study demonstrates that CPT, L-COT SFT, and S-GRPO are all indispensable. CPT provides foundational domain knowledge, L-COT SFT establishes the crucial reasoning-to-code structure, and S-GRPO acts as a refiner that optimizes for security and correctness, leading to a synergistic effect that is essential for overall performance. The parameter sensitivity analysis further confirms that the effectiveness of S-GRPO is highly dependent on a balanced reward configuration.
  };
\end{tikzpicture}
\vspace{5pt} 

3) RQ3: We had human experts rate the reasoning chains of SmartCoder-R1 and two baselines on 150 samples, using a 4-point Likert scale to assess Functionality, Security, and Clarity.

\begin{table*}[ht]
\centering
\caption{Human Evaluation of Reasoning Chain Quality (RQ3). Based on 150 samples, human experts rated the reasoning chains from different models. The scores are on a 4-point Likert scale (1=Poor, 2=Fair, 3=Good, 4=Excellent). The values in the table represent the number of samples receiving each score.}
\label{tab:rq3_human_eval_counts}
\setlength{\tabcolsep}{4pt}
\begin{tabular}{l|l|cccc}
\toprule
\textbf{Model} & \textbf{Dimension} & \textbf{1 (Poor)} & \textbf{2 (Fair)} & \textbf{3 (Good)} & \textbf{4 (Exc.)} \\
\midrule
\multirow{3}{*}{DeepSeek-R1} 
& Functionality & 9 & 27 & 73 & 41 \\
& Security      & 21 & 47 & 56 & 26 \\
& Clarity       & 6  & 16 & 79 & 49 \\
\midrule
\multirow{3}{*}{QwQ-32B} 
& Functionality & 17 & 44 & 66 & 23 \\
& Security      & 36 & 53 & 47 & 14 \\
& Clarity       & 11 & 35 & 73 & 31 \\
\midrule
\multirow{3}{*}{\textbf{SmartCoder-R1 (Ours)}} 
& Functionality & \textbf{7} & \textbf{19} & \textbf{56} & \textbf{68} \\
& Security      & \textbf{6} & \textbf{16} & \textbf{56} & \textbf{72} \\
& Clarity       & \textbf{3} & \textbf{11} & \textbf{57} & \textbf{79} \\
\bottomrule
\end{tabular}
\end{table*}

As detailed in Table \ref{tab:rq3_human_eval_counts}, SmartCoder-R1 demonstrates superior performance across all dimensions. In terms of \textbf{Functionality}, SmartCoder-R1 achieves a high-quality rating (scores 3 or 4) in 82.67\% of cases (124 out of 150), significantly outperforming DeepSeek-R1 (76.00\%) and QwQ-32B (59.33\%). This indicates our model's enhanced ability to accurately interpret complex requirements and formulate a correct implementation plan. The most significant advantage is observed in \textbf{Security}. Our model receives a high-quality rating in 85.33\% of samples (128 out of 150), whereas the baselines lag far behind at 54.67\% for DeepSeek-R1 and only 40.67\% for QwQ-32B. This highlights the effectiveness of our training paradigm in instilling a deep understanding of security principles, a critical factor in smart contract generation. Regarding \textbf{Clarity}, SmartCoder-R1 also leads with 90.67\% of its reasoning chains rated as high-quality, making them easier for developers to audit and trust compared to DeepSeek-R1 (85.33\%) and QwQ-32B (69.33\%). These results confirm that the reasoning chains from SmartCoder-R1 are not only more aligned with functional and security requirements but are also clearer for human developers.

\vspace{5pt} 
\noindent\begin{tikzpicture}
\node[draw=black, thick, fill=gray!20, rounded corners, inner sep=10pt, text width=0.92\linewidth] {
\textbf{Answer to RQ3:} The reasoning chains generated by SmartCoder-R1 demonstrate superior quality in functionality, security, and clarity compared to strong reasoning-focused baselines, making them more reliable for secure smart contract generation.
};
\end{tikzpicture}
\vspace{5pt} 

\begin{figure}[!h]
\centerline{\includegraphics[width=0.8\textwidth]{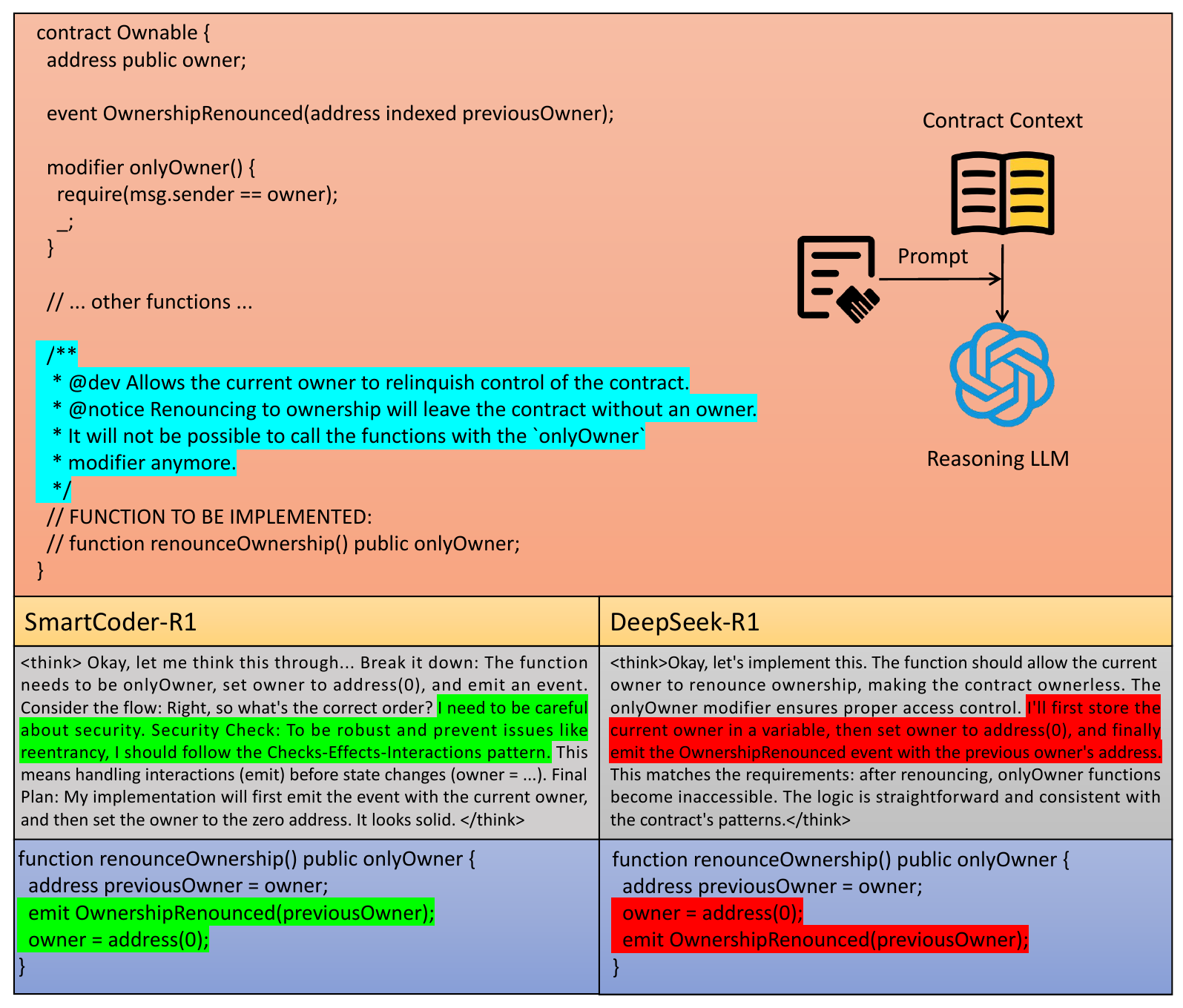}}
\caption{Case study comparing the reasoning process and implementation order of two LLMs (SmartCoder-R1 and DeepSeek-R1) on the renounceOwnership function in smart contracts.}
\label{smart_case}
\end{figure}

4) RQ4: We demonstrate the reasoning and implementation differences between two LLMs in smart contract code generation through a comparative case study. As shown in Fig.~\ref{smart_case}, we examine how SmartCoder-R1 and DeepSeek-R1 (the best baseline) reason about and implement the \texttt{renounceOwnership} function in an \texttt{Ownable} smart contract. In this scenario, both models are tasked with generating a function that allows the contract owner to relinquish control. SmartCoder-R1 not only produces the correct function but also explicitly reasons about security best practices, following the Checks-Effects-Interactions (CEI) pattern to prevent potential reentrancy issues. This is reflected in its implementation order, where the event is emitted before state changes. In contrast, DeepSeek-R1 focuses on functional correctness without considering subtle security concerns, resulting in a direct but potentially less secure implementation where the state change occurs before the event emission. This comparison highlights that while both models can produce functionally correct code, SmartCoder-R1 demonstrates a deeper understanding of security, proactively addressing risks beyond the immediate requirements. DeepSeek-R1, though able to fulfill the prompt, may overlook best practices that are crucial in real-world blockchain environments.

\vspace{5pt} 
\noindent\begin{tikzpicture}
  \node[draw=black, thick, fill=gray!20, rounded corners, inner sep=10pt, text width=0.9\linewidth] {
    \textbf{Answer to RQ4:} SmartCoder-R1 applies security best practices like the CEI pattern, reducing vulnerability risks. In contrast, DeepSeek-R1 gives a solution without security awareness, underscoring the value of security-focused reasoning.
  };
\end{tikzpicture}

\begin{table}[htbp]
\centering
\caption{Key Security Issues in Smart Contract Code Generated by SmartCoder-R1.}
\label{tab:security_issues}
\begin{tabular}{lccc}
\toprule
Security Issue & Count & Percentage (\%) & Risk \\
\midrule
Reentrancy Vulnerabilities & 48 & 32.65 & High \\
Array Bounds Unchecked & 21 & 14.29 & Med \\
Access Control Missing & 18 & 12.24 & High \\
State Validation Missing & 15 & 10.20 & Med \\
Integer Overflow/Underflow & 13 & 8.84 & Med \\
Improper Error Handling & 10 & 6.80 & Med \\
Timestamp Dependence & 6 & 4.08 & Med \\
Gas Limit DoS Risk & 5 & 3.40 & Low \\
Function Visibility Issues & 4 & 2.72 & Low \\
tx.origin Authentication & 3 & 2.04 & High \\
Selfdestruct Usage & 2 & 1.36 & High \\
Delegatecall Context Risk & 2 & 1.36 & High \\
\midrule
\textbf{Total} & \textbf{147} & \textbf{100.00} & -- \\
\bottomrule
\end{tabular}
\end{table}

5) RQ5: To answer RQ5, we conducted an analysis of the residual vulnerabilities in contracts generated by SmartCoder-R1. As shown in Table \ref{tab:security_issues}, our analysis of the 756 test samples revealed that only 83 contained security vulnerabilities, amounting to 147 specific issues. Reentrancy vulnerabilities were the most common, accounting for 32.65\% (48 instances), followed by array bounds unchecked (14.29\%) and missing access controls (12.24\%). Our risk classification is based on potential impact: High-risk vulnerabilities can lead to direct financial loss or loss of contract control; Medium-risk vulnerabilities may cause functional anomalies or partial security threats; and Low-risk issues primarily affect usability or cause minor security concerns. These occurrences primarily stem from the model's occasional oversight of subtle security details within highly complex business logic, such as failing to anticipate all interaction risks in intricate state changes, rather than a general failure in security awareness. This granular analysis is invaluable, guiding our future work towards enhancing the training data with more complex security counterexamples, integrating more advanced static analysis tools into the S-GRPO reward mechanism, and refining the model's reasoning framework to eliminate these specific, high-level vulnerabilities at their source.

\vspace{5pt}
\noindent\begin{tikzpicture}
  \node[draw=black, thick, fill=gray!20, rounded corners, inner sep=10pt, text width=0.92\linewidth] {
    \textbf{Answer to RQ5:} Analysis shows remaining vulnerabilities, mainly reentrancy (32.65\%), occur in complex edge cases, not from a general security flaw. Future work will target these cases by enhancing training data with security counterexamples, integrating static analysis tools, and refining the security reasoning framework.
  };
\end{tikzpicture}

\section{Related Work}


\subsection{Smart Contract Auditing}

Smart contract vulnerability detection has progressed from early program analysis techniques, including symbolic execution (Oyente \cite{luu2016making}), pattern matching (Mythril \cite{mueller2017mythril}, SmartCheck \cite{tikhomirov2018smartcheck}), and formal verification (Securify \cite{tsankov2018securify}), to data-driven models such as deep neural networks \cite{qian2020towards, gao2019smartembed} and graph neural networks (GNNs) \cite{zhuang2020smart, liu2021smart}. Recent advances leverage pre-trained models (Peculiar \cite{wu2021peculiar}), prompt-tuning \cite{yu2023pscvfinder}, and cross-modality learning\cite{qian2023cross} to improve generalization. The latest frontier features LLMs; studies have evaluated general LLMs on real-world datasets \cite{chen2023chatgpt, david2023you}, explored their reasoning capabilities \cite{hu2023large}, and developed LLM-based tools such as GPTScan \cite{sun2024gptscan} and SOChecker \cite{chen2024identifying}. Recent LLM-based tools include Ma et al.'s two-stage iAudit \cite{ma2024combining} and the Smart-LLaMA series by Yu et al. \cite{yu2024smart, yu2025smart}, which evolved from using continual pre-training and supervised fine-tuning to incorporating direct preference optimization for enhanced detection and explanation quality.

\subsection{Smart Contract Code Generation}

In recent years, the security of LLMs in automated smart contract generation has attracted extensive attention. Storhaug et al. \cite{storhaug2023efficient} used vulnerability tagging and constrained decoding to reduce security risks. Wang et al. \cite{wang2025codebc} introduced CodeBC, which improves the security and practicality of generated code through a three-stage fine-tuning strategy. Alam et al. \cite{alam2025solgen} combined static analysis with multi-round masked prompting to achieve automated repair of typical vulnerabilities. Although these approaches have advanced smart contract security, most still lack in-depth modeling of explainability and authentic developer preferences during generation. To address this, we propose a three-stage training paradigm named SmartCoder-R1 that integrates continual pre-training, long chain-of-thought supervised fine-tuning, and security-aware group relative policy optimization, further enhancing the security, explainability, and practical usability of smart contract generation.

\subsection{Reinforcement Learning in LLM Reasoning}

Reinforcement learning (RL) is a rapidly advancing frontier for LLM reasoning \citep{wei2022chain, plaat2024reasoning, wang2024rethinking}. A key approach is outcome-supervised RL, where models like DeepSeek-R1 are rewarded for the correctness of the final result, fostering complex reasoning without step-by-step supervision \citep{guo2025deepseek}. It has proven effective across diverse domains, including programming \cite{wei2025swe, fan2025posterior, wang2025codeboost}, finance \citep{liu2025fin}, vision \citep{huang2025vision, xia2025visionary}, user interface automation \citep{lu2025ui, luo2025gui}, 3D spatial reasoning \citep{pan2025metaspatial} and leveraging external tools \citep{jin2025search, jiang2025deepretrieval}. The success of GRPO in generating domain-specific languages, exemplified by SQL-R1's performance on NL2SQL tasks \citep{ma2025sql}, is particularly relevant. Inspired by these advancements, we adapt outcome-supervised RL to generate secure and functional smart contracts.

\section{Threats to Validity}

\textbf{Internal Validity}: The internal validity of our study hinges on the S-GRPO reward mechanism. While our automated rewards for compilation, security, and format compliance are effective for core objectives, they may not fully capture all nuances of code quality, such as conciseness or stylistic elegance. This could lead SmartCoder-R1 to occasionally generate outputs that are functionally and securely correct but overly verbose. While this does not compromise the primary metrics, it represents a potential area for refinement in the reward function design.

\textbf{External Validity}: The generalizability of SmartCoder-R1's performance is subject to two main considerations. First, our training and evaluation datasets, although comprehensive and curated by experts, represent a subset of the vast and evolving smart contract landscape. The model's performance on highly novel or out-of-distribution contract types may vary. 

\section{Conclusion}

In this paper, we introduced SmartCoder-R1, a novel framework designed to generate secure and explainable smart contracts. Our approach uniquely integrates a three-stage pipeline: Continual Pre-Training to build domain expertise, Long Chain-of-Thought Supervised Fine-Tuning to instill structured security reasoning, and Security-Aware Group Relative Policy Optimization to align the model with security standards. Extensive experiments demonstrate that SmartCoder-R1 significantly outperforms state-of-the-art baselines in generating code that is not only functionally correct and compilable but also substantially more secure and interpretable. 



\bibliographystyle{ACM-Reference-Format}
\bibliography{References}


\begin{thebibliography}{65}


\ifx \showCODEN    \undefined \def \showCODEN     #1{\unskip}     \fi
\ifx \showDOI      \undefined \def \showDOI       #1{#1}\fi
\ifx \showISBNx    \undefined \def \showISBNx     #1{\unskip}     \fi
\ifx \showISBNxiii \undefined \def \showISBNxiii  #1{\unskip}     \fi
\ifx \showISSN     \undefined \def \showISSN      #1{\unskip}     \fi
\ifx \showLCCN     \undefined \def \showLCCN      #1{\unskip}     \fi
\ifx \shownote     \undefined \def \shownote      #1{#1}          \fi
\ifx \showarticletitle \undefined \def \showarticletitle #1{#1}   \fi
\ifx \showURL      \undefined \def \showURL       {\relax}        \fi
\providecommand\bibfield[2]{#2}
\providecommand\bibinfo[2]{#2}
\providecommand\natexlab[1]{#1}
\providecommand\showeprint[2][]{arXiv:#2}

\bibitem[Alam et~al\mbox{.}(2025)]%
        {alam2025solgen}
\bibfield{author}{\bibinfo{person}{Md~Tauseef Alam}, \bibinfo{person}{Sorbajit Goswami}, \bibinfo{person}{Khushi Singh}, \bibinfo{person}{Raju Halder}, \bibinfo{person}{Abyayananda Maiti}, {and} \bibinfo{person}{Soumyadip Banerjee}.} \bibinfo{year}{2025}\natexlab{}.
\newblock \showarticletitle{SolGen: Secure Smart Contract Code Generation Using Large Language Models Via Masked Prompting}. In \bibinfo{booktitle}{\emph{Proceedings of the 18th Innovations in Software Engineering Conference}}. \bibinfo{pages}{1--11}.
\newblock


\bibitem[Alharby and Van~Moorsel(2017)]%
        {alharby2017blockchain}
\bibfield{author}{\bibinfo{person}{Maher Alharby} {and} \bibinfo{person}{Aad Van~Moorsel}.} \bibinfo{year}{2017}\natexlab{}.
\newblock \showarticletitle{Blockchain-based smart contracts: A systematic mapping study}.
\newblock \bibinfo{journal}{\emph{arXiv preprint arXiv:1710.06372}} (\bibinfo{year}{2017}).
\newblock


\bibitem[Chen et~al\mbox{.}(2023)]%
        {chen2023chatgpt}
\bibfield{author}{\bibinfo{person}{Chong Chen}, \bibinfo{person}{Jianzhong Su}, \bibinfo{person}{Jiachi Chen}, \bibinfo{person}{Yanlin Wang}, \bibinfo{person}{Tingting Bi}, \bibinfo{person}{Yanli Wang}, \bibinfo{person}{Xingwei Lin}, \bibinfo{person}{Ting Chen}, {and} \bibinfo{person}{Zibin Zheng}.} \bibinfo{year}{2023}\natexlab{}.
\newblock \showarticletitle{When chatgpt meets smart contract vulnerability detection: How far are we?}
\newblock \bibinfo{journal}{\emph{arXiv preprint arXiv:2309.05520}} (\bibinfo{year}{2023}).
\newblock


\bibitem[Chen et~al\mbox{.}(2024)]%
        {chen2024identifying}
\bibfield{author}{\bibinfo{person}{Jiachi Chen}, \bibinfo{person}{Chong Chen}, \bibinfo{person}{Jiang Hu}, \bibinfo{person}{John Grundy}, \bibinfo{person}{Yanlin Wang}, \bibinfo{person}{Ting Chen}, {and} \bibinfo{person}{Zibin Zheng}.} \bibinfo{year}{2024}\natexlab{}.
\newblock \showarticletitle{Identifying Smart Contract Security Issues in Code Snippets from Stack Overflow}. In \bibinfo{booktitle}{\emph{Proceedings of the 33rd ACM SIGSOFT International Symposium on Software Testing and Analysis}}. \bibinfo{pages}{1198--1210}.
\newblock


\bibitem[David et~al\mbox{.}(2023)]%
        {david2023you}
\bibfield{author}{\bibinfo{person}{Isaac David}, \bibinfo{person}{Liyi Zhou}, \bibinfo{person}{Kaihua Qin}, \bibinfo{person}{Dawn Song}, \bibinfo{person}{Lorenzo Cavallaro}, {and} \bibinfo{person}{Arthur Gervais}.} \bibinfo{year}{2023}\natexlab{}.
\newblock \showarticletitle{Do you still need a manual smart contract audit?}
\newblock \bibinfo{journal}{\emph{arXiv preprint arXiv:2306.12338}} (\bibinfo{year}{2023}).
\newblock


\bibitem[Dhillon et~al\mbox{.}(2017)]%
        {dhillon2017dao}
\bibfield{author}{\bibinfo{person}{Vikram Dhillon}, \bibinfo{person}{David Metcalf}, \bibinfo{person}{Max Hooper}, \bibinfo{person}{Vikram Dhillon}, \bibinfo{person}{David Metcalf}, {and} \bibinfo{person}{Max Hooper}.} \bibinfo{year}{2017}\natexlab{}.
\newblock \showarticletitle{The DAO hacked}.
\newblock \bibinfo{journal}{\emph{blockchain enabled applications: Understand the blockchain Ecosystem and How to Make it work for you}} (\bibinfo{year}{2017}), \bibinfo{pages}{67--78}.
\newblock


\bibitem[Fan et~al\mbox{.}(2025)]%
        {fan2025posterior}
\bibfield{author}{\bibinfo{person}{Lishui Fan}, \bibinfo{person}{Yu Zhang}, \bibinfo{person}{Mouxiang Chen}, {and} \bibinfo{person}{Zhongxin Liu}.} \bibinfo{year}{2025}\natexlab{}.
\newblock \showarticletitle{Posterior-GRPO: Rewarding Reasoning Processes in Code Generation}.
\newblock \bibinfo{journal}{\emph{arXiv preprint arXiv:2508.05170}} (\bibinfo{year}{2025}).
\newblock


\bibitem[Gao et~al\mbox{.}(2019)]%
        {gao2019smartembed}
\bibfield{author}{\bibinfo{person}{Zhipeng Gao}, \bibinfo{person}{Vinoj Jayasundara}, \bibinfo{person}{Lingxiao Jiang}, \bibinfo{person}{Xin Xia}, \bibinfo{person}{David Lo}, {and} \bibinfo{person}{John Grundy}.} \bibinfo{year}{2019}\natexlab{}.
\newblock \showarticletitle{Smartembed: A tool for clone and bug detection in smart contracts through structural code embedding}. In \bibinfo{booktitle}{\emph{2019 IEEE International Conference on Software Maintenance and Evolution (ICSME)}}. IEEE, \bibinfo{pages}{394--397}.
\newblock


\bibitem[Guo et~al\mbox{.}(2025)]%
        {guo2025deepseek}
\bibfield{author}{\bibinfo{person}{Daya Guo}, \bibinfo{person}{Dejian Yang}, \bibinfo{person}{Haowei Zhang}, \bibinfo{person}{Junxiao Song}, \bibinfo{person}{Ruoyu Zhang}, \bibinfo{person}{Runxin Xu}, \bibinfo{person}{Qihao Zhu}, \bibinfo{person}{Shirong Ma}, \bibinfo{person}{Peiyi Wang}, \bibinfo{person}{Xiao Bi}, {et~al\mbox{.}}} \bibinfo{year}{2025}\natexlab{}.
\newblock \showarticletitle{Deepseek-r1: Incentivizing reasoning capability in llms via reinforcement learning}.
\newblock \bibinfo{journal}{\emph{arXiv preprint arXiv:2501.12948}} (\bibinfo{year}{2025}).
\newblock


\bibitem[Guo et~al\mbox{.}(2024)]%
        {guo2024deepseek}
\bibfield{author}{\bibinfo{person}{Daya Guo}, \bibinfo{person}{Qihao Zhu}, \bibinfo{person}{Dejian Yang}, \bibinfo{person}{Zhenda Xie}, \bibinfo{person}{Kai Dong}, \bibinfo{person}{Wentao Zhang}, \bibinfo{person}{Guanting Chen}, \bibinfo{person}{Xiao Bi}, \bibinfo{person}{Yu Wu}, \bibinfo{person}{YK Li}, {et~al\mbox{.}}} \bibinfo{year}{2024}\natexlab{}.
\newblock \showarticletitle{DeepSeek-Coder: When the Large Language Model Meets Programming--The Rise of Code Intelligence}.
\newblock \bibinfo{journal}{\emph{arXiv preprint arXiv:2401.14196}} (\bibinfo{year}{2024}).
\newblock


\bibitem[Heged{\H{u}}s(2018)]%
        {hegedHus2018towards}
\bibfield{author}{\bibinfo{person}{P{\'e}ter Heged{\H{u}}s}.} \bibinfo{year}{2018}\natexlab{}.
\newblock \showarticletitle{Towards analyzing the complexity landscape of solidity based ethereum smart contracts}. In \bibinfo{booktitle}{\emph{Proceedings of the 1st International Workshop on Emerging Trends in Software Engineering for Blockchain}}. \bibinfo{pages}{35--39}.
\newblock


\bibitem[Hewa et~al\mbox{.}(2021)]%
        {hewa2021survey}
\bibfield{author}{\bibinfo{person}{Tharaka Hewa}, \bibinfo{person}{Mika Ylianttila}, {and} \bibinfo{person}{Madhusanka Liyanage}.} \bibinfo{year}{2021}\natexlab{}.
\newblock \showarticletitle{Survey on blockchain based smart contracts: Applications, opportunities and challenges}.
\newblock \bibinfo{journal}{\emph{Journal of Network and Computer Applications}}  \bibinfo{volume}{177} (\bibinfo{year}{2021}), \bibinfo{pages}{102857}.
\newblock


\bibitem[Hu et~al\mbox{.}(2023)]%
        {hu2023large}
\bibfield{author}{\bibinfo{person}{Sihao Hu}, \bibinfo{person}{Tiansheng Huang}, \bibinfo{person}{Fatih {\.I}lhan}, \bibinfo{person}{Selim~Furkan Tekin}, {and} \bibinfo{person}{Ling Liu}.} \bibinfo{year}{2023}\natexlab{}.
\newblock \showarticletitle{Large language model-powered smart contract vulnerability detection: New perspectives}.
\newblock \bibinfo{journal}{\emph{arXiv preprint arXiv:2310.01152}} (\bibinfo{year}{2023}).
\newblock


\bibitem[Huang et~al\mbox{.}(2025)]%
        {huang2025vision}
\bibfield{author}{\bibinfo{person}{Wenxuan Huang}, \bibinfo{person}{Bohan Jia}, \bibinfo{person}{Zijie Zhai}, \bibinfo{person}{Shaosheng Cao}, \bibinfo{person}{Zheyu Ye}, \bibinfo{person}{Fei Zhao}, \bibinfo{person}{Zhe Xu}, \bibinfo{person}{Yao Hu}, {and} \bibinfo{person}{Shaohui Lin}.} \bibinfo{year}{2025}\natexlab{}.
\newblock \showarticletitle{Vision-r1: Incentivizing reasoning capability in multimodal large language models}.
\newblock \bibinfo{journal}{\emph{arXiv preprint arXiv:2503.06749}} (\bibinfo{year}{2025}).
\newblock


\bibitem[Hui et~al\mbox{.}(2024)]%
        {hui2024qwen2}
\bibfield{author}{\bibinfo{person}{Binyuan Hui}, \bibinfo{person}{Jian Yang}, \bibinfo{person}{Zeyu Cui}, \bibinfo{person}{Jiaxi Yang}, \bibinfo{person}{Dayiheng Liu}, \bibinfo{person}{Lei Zhang}, \bibinfo{person}{Tianyu Liu}, \bibinfo{person}{Jiajun Zhang}, \bibinfo{person}{Bowen Yu}, \bibinfo{person}{Keming Lu}, {et~al\mbox{.}}} \bibinfo{year}{2024}\natexlab{}.
\newblock \showarticletitle{Qwen2. 5-coder technical report}.
\newblock \bibinfo{journal}{\emph{arXiv preprint arXiv:2409.12186}} (\bibinfo{year}{2024}).
\newblock


\bibitem[Jiang et~al\mbox{.}(2025)]%
        {jiang2025deepretrieval}
\bibfield{author}{\bibinfo{person}{Pengcheng Jiang}, \bibinfo{person}{Jiacheng Lin}, \bibinfo{person}{Lang Cao}, \bibinfo{person}{Runchu Tian}, \bibinfo{person}{SeongKu Kang}, \bibinfo{person}{Zifeng Wang}, \bibinfo{person}{Jimeng Sun}, {and} \bibinfo{person}{Jiawei Han}.} \bibinfo{year}{2025}\natexlab{}.
\newblock \showarticletitle{Deepretrieval: Hacking real search engines and retrievers with large language models via reinforcement learning}.
\newblock \bibinfo{journal}{\emph{arXiv preprint arXiv:2503.00223}} (\bibinfo{year}{2025}).
\newblock


\bibitem[Jin et~al\mbox{.}(2025)]%
        {jin2025search}
\bibfield{author}{\bibinfo{person}{Bowen Jin}, \bibinfo{person}{Hansi Zeng}, \bibinfo{person}{Zhenrui Yue}, \bibinfo{person}{Jinsung Yoon}, \bibinfo{person}{Sercan Arik}, \bibinfo{person}{Dong Wang}, \bibinfo{person}{Hamed Zamani}, {and} \bibinfo{person}{Jiawei Han}.} \bibinfo{year}{2025}\natexlab{}.
\newblock \showarticletitle{Search-r1: Training llms to reason and leverage search engines with reinforcement learning}.
\newblock \bibinfo{journal}{\emph{arXiv preprint arXiv:2503.09516}} (\bibinfo{year}{2025}).
\newblock


\bibitem[Joshi et~al\mbox{.}(2015)]%
        {joshi2015likert}
\bibfield{author}{\bibinfo{person}{Ankur Joshi}, \bibinfo{person}{Saket Kale}, \bibinfo{person}{Satish Chandel}, {and} \bibinfo{person}{D~Kumar Pal}.} \bibinfo{year}{2015}\natexlab{}.
\newblock \showarticletitle{Likert scale: Explored and explained}.
\newblock \bibinfo{journal}{\emph{British journal of applied science \& technology}} \bibinfo{volume}{7}, \bibinfo{number}{4} (\bibinfo{year}{2015}), \bibinfo{pages}{396--403}.
\newblock


\bibitem[Liu et~al\mbox{.}(2024)]%
        {liu2024deepseek}
\bibfield{author}{\bibinfo{person}{Aixin Liu}, \bibinfo{person}{Bei Feng}, \bibinfo{person}{Bing Xue}, \bibinfo{person}{Bingxuan Wang}, \bibinfo{person}{Bochao Wu}, \bibinfo{person}{Chengda Lu}, \bibinfo{person}{Chenggang Zhao}, \bibinfo{person}{Chengqi Deng}, \bibinfo{person}{Chenyu Zhang}, \bibinfo{person}{Chong Ruan}, {et~al\mbox{.}}} \bibinfo{year}{2024}\natexlab{}.
\newblock \showarticletitle{Deepseek-v3 technical report}.
\newblock \bibinfo{journal}{\emph{arXiv preprint arXiv:2412.19437}} (\bibinfo{year}{2024}).
\newblock


\bibitem[Liu et~al\mbox{.}(2025)]%
        {liu2025fin}
\bibfield{author}{\bibinfo{person}{Zhaowei Liu}, \bibinfo{person}{Xin Guo}, \bibinfo{person}{Fangqi Lou}, \bibinfo{person}{Lingfeng Zeng}, \bibinfo{person}{Jinyi Niu}, \bibinfo{person}{Zixuan Wang}, \bibinfo{person}{Jiajie Xu}, \bibinfo{person}{Weige Cai}, \bibinfo{person}{Ziwei Yang}, \bibinfo{person}{Xueqian Zhao}, {et~al\mbox{.}}} \bibinfo{year}{2025}\natexlab{}.
\newblock \showarticletitle{Fin-r1: A large language model for financial reasoning through reinforcement learning}.
\newblock \bibinfo{journal}{\emph{arXiv preprint arXiv:2503.16252}} (\bibinfo{year}{2025}).
\newblock


\bibitem[Liu et~al\mbox{.}(2021)]%
        {liu2021smart}
\bibfield{author}{\bibinfo{person}{Zhenguang Liu}, \bibinfo{person}{Peng Qian}, \bibinfo{person}{Xiang Wang}, \bibinfo{person}{Lei Zhu}, \bibinfo{person}{Qinming He}, {and} \bibinfo{person}{Shouling Ji}.} \bibinfo{year}{2021}\natexlab{}.
\newblock \showarticletitle{Smart contract vulnerability detection: from pure neural network to interpretable graph feature and expert pattern fusion}.
\newblock \bibinfo{journal}{\emph{arXiv preprint arXiv:2106.09282}} (\bibinfo{year}{2021}).
\newblock


\bibitem[Loshchilov and Hutter(2017)]%
        {adamw}
\bibfield{author}{\bibinfo{person}{Ilya Loshchilov} {and} \bibinfo{person}{Frank Hutter}.} \bibinfo{year}{2017}\natexlab{}.
\newblock \showarticletitle{Decoupled weight decay regularization}.
\newblock \bibinfo{journal}{\emph{arXiv preprint arXiv:1711.05101}} (\bibinfo{year}{2017}).
\newblock


\bibitem[Lu et~al\mbox{.}(2025)]%
        {lu2025ui}
\bibfield{author}{\bibinfo{person}{Zhengxi Lu}, \bibinfo{person}{Yuxiang Chai}, \bibinfo{person}{Yaxuan Guo}, \bibinfo{person}{Xi Yin}, \bibinfo{person}{Liang Liu}, \bibinfo{person}{Hao Wang}, \bibinfo{person}{Han Xiao}, \bibinfo{person}{Shuai Ren}, \bibinfo{person}{Guanjing Xiong}, {and} \bibinfo{person}{Hongsheng Li}.} \bibinfo{year}{2025}\natexlab{}.
\newblock \showarticletitle{UI-R1: Enhancing Efficient Action Prediction of GUI Agents by Reinforcement Learning}.
\newblock \bibinfo{journal}{\emph{arXiv preprint arXiv:2503.21620}} (\bibinfo{year}{2025}).
\newblock


\bibitem[Luo et~al\mbox{.}(2025)]%
        {luo2025gui}
\bibfield{author}{\bibinfo{person}{Run Luo}, \bibinfo{person}{Lu Wang}, \bibinfo{person}{Wanwei He}, {and} \bibinfo{person}{Xiaobo Xia}.} \bibinfo{year}{2025}\natexlab{}.
\newblock \showarticletitle{Gui-r1: A generalist r1-style vision-language action model for gui agents}.
\newblock \bibinfo{journal}{\emph{arXiv preprint arXiv:2504.10458}} (\bibinfo{year}{2025}).
\newblock


\bibitem[Luu et~al\mbox{.}(2016)]%
        {luu2016making}
\bibfield{author}{\bibinfo{person}{Loi Luu}, \bibinfo{person}{Duc-Hiep Chu}, \bibinfo{person}{Hrishi Olickel}, \bibinfo{person}{Prateek Saxena}, {and} \bibinfo{person}{Aquinas Hobor}.} \bibinfo{year}{2016}\natexlab{}.
\newblock \showarticletitle{Making smart contracts smarter}. In \bibinfo{booktitle}{\emph{Proceedings of the 2016 ACM SIGSAC conference on computer and communications security}}. \bibinfo{pages}{254--269}.
\newblock


\bibitem[Ma et~al\mbox{.}(2025)]%
        {ma2025sql}
\bibfield{author}{\bibinfo{person}{Peixian Ma}, \bibinfo{person}{Xialie Zhuang}, \bibinfo{person}{Chengjin Xu}, \bibinfo{person}{Xuhui Jiang}, \bibinfo{person}{Ran Chen}, {and} \bibinfo{person}{Jian Guo}.} \bibinfo{year}{2025}\natexlab{}.
\newblock \showarticletitle{Sql-r1: Training natural language to sql reasoning model by reinforcement learning}.
\newblock \bibinfo{journal}{\emph{arXiv preprint arXiv:2504.08600}} (\bibinfo{year}{2025}).
\newblock


\bibitem[Ma et~al\mbox{.}(2024)]%
        {ma2024combining}
\bibfield{author}{\bibinfo{person}{Wei Ma}, \bibinfo{person}{Daoyuan Wu}, \bibinfo{person}{Yuqiang Sun}, \bibinfo{person}{Tianwen Wang}, \bibinfo{person}{Shangqing Liu}, \bibinfo{person}{Jian Zhang}, \bibinfo{person}{Yue Xue}, {and} \bibinfo{person}{Yang Liu}.} \bibinfo{year}{2024}\natexlab{}.
\newblock \showarticletitle{Combining Fine-Tuning and LLM-based Agents for Intuitive Smart Contract Auditing with Justifications}.
\newblock \bibinfo{journal}{\emph{arXiv preprint arXiv:2403.16073}} (\bibinfo{year}{2024}).
\newblock


\bibitem[Mehar et~al\mbox{.}(2019)]%
        {mehar2019understanding}
\bibfield{author}{\bibinfo{person}{Muhammad~Izhar Mehar}, \bibinfo{person}{Charles~Louis Shier}, \bibinfo{person}{Alana Giambattista}, \bibinfo{person}{Elgar Gong}, \bibinfo{person}{Gabrielle Fletcher}, \bibinfo{person}{Ryan Sanayhie}, \bibinfo{person}{Henry~M Kim}, {and} \bibinfo{person}{Marek Laskowski}.} \bibinfo{year}{2019}\natexlab{}.
\newblock \showarticletitle{Understanding a revolutionary and flawed grand experiment in blockchain: the DAO attack}.
\newblock \bibinfo{journal}{\emph{Journal of Cases on Information Technology (JCIT)}} \bibinfo{volume}{21}, \bibinfo{number}{1} (\bibinfo{year}{2019}), \bibinfo{pages}{19--32}.
\newblock


\bibitem[Mueller(2017)]%
        {mueller2017mythril}
\bibfield{author}{\bibinfo{person}{B Mueller}.} \bibinfo{year}{2017}\natexlab{}.
\newblock \bibinfo{title}{Mythril-Reversing and bug hunting framework for the Ethereum blockchain}.
\newblock
\newblock


\bibitem[OpenAI(2025)]%
        {openai_gpt41_2025}
\bibfield{author}{\bibinfo{person}{OpenAI}.} \bibinfo{year}{2025}\natexlab{}.
\newblock \bibinfo{title}{GPT-4.1}.
\newblock \bibinfo{howpublished}{\url{https://platform.openai.com/docs/models/gpt-4.1}}.
\newblock


\bibitem[Pan and Liu(2025)]%
        {pan2025metaspatial}
\bibfield{author}{\bibinfo{person}{Zhenyu Pan} {and} \bibinfo{person}{Han Liu}.} \bibinfo{year}{2025}\natexlab{}.
\newblock \showarticletitle{Metaspatial: Reinforcing 3d spatial reasoning in vlms for the metaverse}.
\newblock \bibinfo{journal}{\emph{arXiv preprint arXiv:2503.18470}} (\bibinfo{year}{2025}).
\newblock


\bibitem[Papineni et~al\mbox{.}(2002)]%
        {papineni2002bleu}
\bibfield{author}{\bibinfo{person}{Kishore Papineni}, \bibinfo{person}{Salim Roukos}, \bibinfo{person}{Todd Ward}, {and} \bibinfo{person}{Wei-Jing Zhu}.} \bibinfo{year}{2002}\natexlab{}.
\newblock \showarticletitle{Bleu: a method for automatic evaluation of machine translation}. In \bibinfo{booktitle}{\emph{Proceedings of the 40th annual meeting of the Association for Computational Linguistics}}. \bibinfo{pages}{311--318}.
\newblock


\bibitem[Plaat et~al\mbox{.}(2024)]%
        {plaat2024reasoning}
\bibfield{author}{\bibinfo{person}{Aske Plaat}, \bibinfo{person}{Annie Wong}, \bibinfo{person}{Suzan Verberne}, \bibinfo{person}{Joost Broekens}, \bibinfo{person}{Niki van Stein}, {and} \bibinfo{person}{Thomas Back}.} \bibinfo{year}{2024}\natexlab{}.
\newblock \showarticletitle{Reasoning with large language models, a survey}.
\newblock \bibinfo{journal}{\emph{arXiv preprint arXiv:2407.11511}} (\bibinfo{year}{2024}).
\newblock


\bibitem[Qian et~al\mbox{.}(2020)]%
        {qian2020towards}
\bibfield{author}{\bibinfo{person}{Peng Qian}, \bibinfo{person}{Zhenguang Liu}, \bibinfo{person}{Qinming He}, \bibinfo{person}{Roger Zimmermann}, {and} \bibinfo{person}{Xun Wang}.} \bibinfo{year}{2020}\natexlab{}.
\newblock \showarticletitle{Towards automated reentrancy detection for smart contracts based on sequential models}.
\newblock \bibinfo{journal}{\emph{IEEE Access}}  \bibinfo{volume}{8} (\bibinfo{year}{2020}), \bibinfo{pages}{19685--19695}.
\newblock


\bibitem[Qian et~al\mbox{.}(2023)]%
        {qian2023cross}
\bibfield{author}{\bibinfo{person}{Peng Qian}, \bibinfo{person}{Zhenguang Liu}, \bibinfo{person}{Yifang Yin}, {and} \bibinfo{person}{Qinming He}.} \bibinfo{year}{2023}\natexlab{}.
\newblock \showarticletitle{Cross-modality mutual learning for enhancing smart contract vulnerability detection on bytecode}. In \bibinfo{booktitle}{\emph{Proceedings of the ACM Web Conference 2023}}. \bibinfo{pages}{2220--2229}.
\newblock


\bibitem[Qwen(2025)]%
        {qwen_qwq_2025}
\bibfield{author}{\bibinfo{person}{Qwen}.} \bibinfo{year}{2025}\natexlab{}.
\newblock \bibinfo{title}{QwQ-32B}.
\newblock \bibinfo{howpublished}{\url{https://huggingface.co/Qwen/QwQ-32B}}.
\newblock


\bibitem[Rasley et~al\mbox{.}(2020)]%
        {rasley2020deepspeed}
\bibfield{author}{\bibinfo{person}{Jeff Rasley}, \bibinfo{person}{Samyam Rajbhandari}, \bibinfo{person}{Olatunji Ruwase}, {and} \bibinfo{person}{Yuxiong He}.} \bibinfo{year}{2020}\natexlab{}.
\newblock \showarticletitle{Deepspeed: System optimizations enable training deep learning models with over 100 billion parameters}. In \bibinfo{booktitle}{\emph{Proceedings of the 26th ACM SIGKDD International Conference on Knowledge Discovery \& Data Mining}}. \bibinfo{pages}{3505--3506}.
\newblock


\bibitem[Ren et~al\mbox{.}(2020)]%
        {ren2020codebleu}
\bibfield{author}{\bibinfo{person}{Shuo Ren}, \bibinfo{person}{Daya Guo}, \bibinfo{person}{Shuai Lu}, \bibinfo{person}{Long Zhou}, \bibinfo{person}{Shujie Liu}, \bibinfo{person}{Duyu Tang}, \bibinfo{person}{Neel Sundaresan}, \bibinfo{person}{Ming Zhou}, \bibinfo{person}{Ambrosio Blanco}, {and} \bibinfo{person}{Shuai Ma}.} \bibinfo{year}{2020}\natexlab{}.
\newblock \showarticletitle{Codebleu: a method for automatic evaluation of code synthesis}.
\newblock \bibinfo{journal}{\emph{arXiv preprint arXiv:2009.10297}} (\bibinfo{year}{2020}).
\newblock


\bibitem[Roziere et~al\mbox{.}(2023)]%
        {roziere2023code}
\bibfield{author}{\bibinfo{person}{Baptiste Roziere}, \bibinfo{person}{Jonas Gehring}, \bibinfo{person}{Fabian Gloeckle}, \bibinfo{person}{Sten Sootla}, \bibinfo{person}{Itai Gat}, \bibinfo{person}{Xiaoqing~Ellen Tan}, \bibinfo{person}{Yossi Adi}, \bibinfo{person}{Jingyu Liu}, \bibinfo{person}{Romain Sauvestre}, \bibinfo{person}{Tal Remez}, {et~al\mbox{.}}} \bibinfo{year}{2023}\natexlab{}.
\newblock \showarticletitle{Code llama: Open foundation models for code}.
\newblock \bibinfo{journal}{\emph{arXiv preprint arXiv:2308.12950}} (\bibinfo{year}{2023}).
\newblock


\bibitem[Sheng et~al\mbox{.}(2025)]%
        {sheng2025hybridflow}
\bibfield{author}{\bibinfo{person}{Guangming Sheng}, \bibinfo{person}{Chi Zhang}, \bibinfo{person}{Zilingfeng Ye}, \bibinfo{person}{Xibin Wu}, \bibinfo{person}{Wang Zhang}, \bibinfo{person}{Ru Zhang}, \bibinfo{person}{Yanghua Peng}, \bibinfo{person}{Haibin Lin}, {and} \bibinfo{person}{Chuan Wu}.} \bibinfo{year}{2025}\natexlab{}.
\newblock \showarticletitle{Hybridflow: A flexible and efficient rlhf framework}. In \bibinfo{booktitle}{\emph{Proceedings of the Twentieth European Conference on Computer Systems}}. \bibinfo{pages}{1279--1297}.
\newblock


\bibitem[Storhaug et~al\mbox{.}(2023)]%
        {storhaug2023efficient}
\bibfield{author}{\bibinfo{person}{Andr{\'e} Storhaug}, \bibinfo{person}{Jingyue Li}, {and} \bibinfo{person}{Tianyuan Hu}.} \bibinfo{year}{2023}\natexlab{}.
\newblock \showarticletitle{Efficient avoidance of vulnerabilities in auto-completed smart contract code using vulnerability-constrained decoding}. In \bibinfo{booktitle}{\emph{2023 IEEE 34th International Symposium on Software Reliability Engineering (ISSRE)}}. IEEE, \bibinfo{pages}{683--693}.
\newblock


\bibitem[Sun et~al\mbox{.}(2024)]%
        {sun2024gptscan}
\bibfield{author}{\bibinfo{person}{Yuqiang Sun}, \bibinfo{person}{Daoyuan Wu}, \bibinfo{person}{Yue Xue}, \bibinfo{person}{Han Liu}, \bibinfo{person}{Haijun Wang}, \bibinfo{person}{Zhengzi Xu}, \bibinfo{person}{Xiaofei Xie}, {and} \bibinfo{person}{Yang Liu}.} \bibinfo{year}{2024}\natexlab{}.
\newblock \showarticletitle{GPTScan: Detecting Logic Vulnerabilities in Smart Contracts by Combining GPT with Program Analysis}.
\newblock \bibinfo{journal}{\emph{Proc. IEEE/ACM ICSE}} (\bibinfo{year}{2024}).
\newblock


\bibitem[Swan(2015)]%
        {swan2015blockchain}
\bibfield{author}{\bibinfo{person}{Melanie Swan}.} \bibinfo{year}{2015}\natexlab{}.
\newblock \bibinfo{booktitle}{\emph{Blockchain: Blueprint for a new economy}}.
\newblock \bibinfo{publisher}{" O'Reilly Media, Inc."}.
\newblock


\bibitem[Team(2024)]%
        {team2024qwen2}
\bibfield{author}{\bibinfo{person}{Qwen Team}.} \bibinfo{year}{2024}\natexlab{}.
\newblock \showarticletitle{Qwen2 technical report}.
\newblock \bibinfo{journal}{\emph{arXiv preprint arXiv:2407.10671}} (\bibinfo{year}{2024}).
\newblock


\bibitem[Tikhomirov et~al\mbox{.}(2018)]%
        {tikhomirov2018smartcheck}
\bibfield{author}{\bibinfo{person}{Sergei Tikhomirov}, \bibinfo{person}{Ekaterina Voskresenskaya}, \bibinfo{person}{Ivan Ivanitskiy}, \bibinfo{person}{Ramil Takhaviev}, \bibinfo{person}{Evgeny Marchenko}, {and} \bibinfo{person}{Yaroslav Alexandrov}.} \bibinfo{year}{2018}\natexlab{}.
\newblock \showarticletitle{Smartcheck: Static analysis of ethereum smart contracts}. In \bibinfo{booktitle}{\emph{Proceedings of the 1st International Workshop on Emerging Trends in Software Engineering for Blockchain}}. \bibinfo{pages}{9--16}.
\newblock


\bibitem[Tsankov et~al\mbox{.}(2018)]%
        {tsankov2018securify}
\bibfield{author}{\bibinfo{person}{Petar Tsankov}, \bibinfo{person}{Andrei Dan}, \bibinfo{person}{Dana Drachsler-Cohen}, \bibinfo{person}{Arthur Gervais}, \bibinfo{person}{Florian Buenzli}, {and} \bibinfo{person}{Martin Vechev}.} \bibinfo{year}{2018}\natexlab{}.
\newblock \showarticletitle{Securify: Practical security analysis of smart contracts}. In \bibinfo{booktitle}{\emph{Proceedings of the 2018 ACM SIGSAC Conference on Computer and Communications Security}}. \bibinfo{pages}{67--82}.
\newblock


\bibitem[Wang et~al\mbox{.}(2025b)]%
        {wang2025codebc}
\bibfield{author}{\bibinfo{person}{Lingxiang Wang}, \bibinfo{person}{Hainan Zhang}, \bibinfo{person}{Qinnan Zhang}, \bibinfo{person}{Ziwei Wang}, \bibinfo{person}{Hongwei Zheng}, \bibinfo{person}{Jin Dong}, {and} \bibinfo{person}{Zhiming Zheng}.} \bibinfo{year}{2025}\natexlab{b}.
\newblock \showarticletitle{CodeBC: A More Secure Large Language Model for Smart Contract Code Generation in Blockchain}.
\newblock \bibinfo{journal}{\emph{arXiv preprint arXiv:2504.21043}} (\bibinfo{year}{2025}).
\newblock


\bibitem[Wang et~al\mbox{.}(2024)]%
        {wang2024rethinking}
\bibfield{author}{\bibinfo{person}{Qineng Wang}, \bibinfo{person}{Zihao Wang}, \bibinfo{person}{Ying Su}, \bibinfo{person}{Hanghang Tong}, {and} \bibinfo{person}{Yangqiu Song}.} \bibinfo{year}{2024}\natexlab{}.
\newblock \showarticletitle{Rethinking the bounds of llm reasoning: Are multi-agent discussions the key?}
\newblock \bibinfo{journal}{\emph{arXiv preprint arXiv:2402.18272}} (\bibinfo{year}{2024}).
\newblock


\bibitem[Wang et~al\mbox{.}(2025a)]%
        {wang2025codeboost}
\bibfield{author}{\bibinfo{person}{Sijie Wang}, \bibinfo{person}{Quanjiang Guo}, \bibinfo{person}{Kai Zhao}, \bibinfo{person}{Yawei Zhang}, \bibinfo{person}{Xin Li}, \bibinfo{person}{Xiang Li}, \bibinfo{person}{Siqi Li}, \bibinfo{person}{Rui She}, \bibinfo{person}{Shangshu Yu}, {and} \bibinfo{person}{Wee~Peng Tay}.} \bibinfo{year}{2025}\natexlab{a}.
\newblock \showarticletitle{CodeBoost: Boosting Code LLMs by Squeezing Knowledge from Code Snippets with RL}.
\newblock \bibinfo{journal}{\emph{arXiv preprint arXiv:2508.05242}} (\bibinfo{year}{2025}).
\newblock


\bibitem[Wang et~al\mbox{.}(2025c)]%
        {wang2025octothinker}
\bibfield{author}{\bibinfo{person}{Zengzhi Wang}, \bibinfo{person}{Fan Zhou}, \bibinfo{person}{Xuefeng Li}, {and} \bibinfo{person}{Pengfei Liu}.} \bibinfo{year}{2025}\natexlab{c}.
\newblock \showarticletitle{Octothinker: Mid-training incentivizes reinforcement learning scaling}.
\newblock \bibinfo{journal}{\emph{arXiv preprint arXiv:2506.20512}} (\bibinfo{year}{2025}).
\newblock


\bibitem[Wei et~al\mbox{.}(2022)]%
        {wei2022chain}
\bibfield{author}{\bibinfo{person}{Jason Wei}, \bibinfo{person}{Xuezhi Wang}, \bibinfo{person}{Dale Schuurmans}, \bibinfo{person}{Maarten Bosma}, \bibinfo{person}{Fei Xia}, \bibinfo{person}{Ed Chi}, \bibinfo{person}{Quoc~V Le}, \bibinfo{person}{Denny Zhou}, {et~al\mbox{.}}} \bibinfo{year}{2022}\natexlab{}.
\newblock \showarticletitle{Chain-of-thought prompting elicits reasoning in large language models}.
\newblock \bibinfo{journal}{\emph{Advances in neural information processing systems}}  \bibinfo{volume}{35} (\bibinfo{year}{2022}), \bibinfo{pages}{24824--24837}.
\newblock


\bibitem[Wei et~al\mbox{.}(2025)]%
        {wei2025swe}
\bibfield{author}{\bibinfo{person}{Yuxiang Wei}, \bibinfo{person}{Olivier Duchenne}, \bibinfo{person}{Jade Copet}, \bibinfo{person}{Quentin Carbonneaux}, \bibinfo{person}{Lingming Zhang}, \bibinfo{person}{Daniel Fried}, \bibinfo{person}{Gabriel Synnaeve}, \bibinfo{person}{Rishabh Singh}, {and} \bibinfo{person}{Sida~I Wang}.} \bibinfo{year}{2025}\natexlab{}.
\newblock \showarticletitle{Swe-rl: Advancing llm reasoning via reinforcement learning on open software evolution}.
\newblock \bibinfo{journal}{\emph{arXiv preprint arXiv:2502.18449}} (\bibinfo{year}{2025}).
\newblock


\bibitem[Wood et~al\mbox{.}(2014)]%
        {wood2014ethereum}
\bibfield{author}{\bibinfo{person}{Gavin Wood} {et~al\mbox{.}}} \bibinfo{year}{2014}\natexlab{}.
\newblock \showarticletitle{Ethereum: A secure decentralised generalised transaction ledger}.
\newblock \bibinfo{journal}{\emph{Ethereum project yellow paper}} \bibinfo{volume}{151}, \bibinfo{number}{2014} (\bibinfo{year}{2014}), \bibinfo{pages}{1--32}.
\newblock


\bibitem[Wu et~al\mbox{.}(2021)]%
        {wu2021peculiar}
\bibfield{author}{\bibinfo{person}{Hongjun Wu}, \bibinfo{person}{Zhuo Zhang}, \bibinfo{person}{Shangwen Wang}, \bibinfo{person}{Yan Lei}, \bibinfo{person}{Bo Lin}, \bibinfo{person}{Yihao Qin}, \bibinfo{person}{Haoyu Zhang}, {and} \bibinfo{person}{Xiaoguang Mao}.} \bibinfo{year}{2021}\natexlab{}.
\newblock \showarticletitle{Peculiar: Smart contract vulnerability detection based on crucial data flow graph and pre-training techniques}. In \bibinfo{booktitle}{\emph{2021 IEEE 32nd International Symposium on Software Reliability Engineering (ISSRE)}}. IEEE, \bibinfo{pages}{378--389}.
\newblock


\bibitem[Xia et~al\mbox{.}(2025)]%
        {xia2025visionary}
\bibfield{author}{\bibinfo{person}{Jiaer Xia}, \bibinfo{person}{Yuhang Zang}, \bibinfo{person}{Peng Gao}, \bibinfo{person}{Yixuan Li}, {and} \bibinfo{person}{Kaiyang Zhou}.} \bibinfo{year}{2025}\natexlab{}.
\newblock \showarticletitle{Visionary-r1: Mitigating shortcuts in visual reasoning with reinforcement learning}.
\newblock \bibinfo{journal}{\emph{arXiv preprint arXiv:2505.14677}} (\bibinfo{year}{2025}).
\newblock


\bibitem[Xie et~al\mbox{.}(2025)]%
        {xie2025logic}
\bibfield{author}{\bibinfo{person}{Tian Xie}, \bibinfo{person}{Zitian Gao}, \bibinfo{person}{Qingnan Ren}, \bibinfo{person}{Haoming Luo}, \bibinfo{person}{Yuqian Hong}, \bibinfo{person}{Bryan Dai}, \bibinfo{person}{Joey Zhou}, \bibinfo{person}{Kai Qiu}, \bibinfo{person}{Zhirong Wu}, {and} \bibinfo{person}{Chong Luo}.} \bibinfo{year}{2025}\natexlab{}.
\newblock \showarticletitle{Logic-rl: Unleashing llm reasoning with rule-based reinforcement learning}.
\newblock \bibinfo{journal}{\emph{arXiv preprint arXiv:2502.14768}} (\bibinfo{year}{2025}).
\newblock


\bibitem[Yang et~al\mbox{.}(2025)]%
        {yang2025qwen3}
\bibfield{author}{\bibinfo{person}{An Yang}, \bibinfo{person}{Anfeng Li}, \bibinfo{person}{Baosong Yang}, \bibinfo{person}{Beichen Zhang}, \bibinfo{person}{Binyuan Hui}, \bibinfo{person}{Bo Zheng}, \bibinfo{person}{Bowen Yu}, \bibinfo{person}{Chang Gao}, \bibinfo{person}{Chengen Huang}, \bibinfo{person}{Chenxu Lv}, {et~al\mbox{.}}} \bibinfo{year}{2025}\natexlab{}.
\newblock \showarticletitle{Qwen3 technical report}.
\newblock \bibinfo{journal}{\emph{arXiv preprint arXiv:2505.09388}} (\bibinfo{year}{2025}).
\newblock


\bibitem[Yu et~al\mbox{.}(2024)]%
        {yu2024smart}
\bibfield{author}{\bibinfo{person}{Lei Yu}, \bibinfo{person}{Shiqi Chen}, \bibinfo{person}{Hang Yuan}, \bibinfo{person}{Peng Wang}, \bibinfo{person}{Zhirong Huang}, \bibinfo{person}{Jingyuan Zhang}, \bibinfo{person}{Chenjie Shen}, \bibinfo{person}{Fengjun Zhang}, \bibinfo{person}{Li Yang}, {and} \bibinfo{person}{Jiajia Ma}.} \bibinfo{year}{2024}\natexlab{}.
\newblock \showarticletitle{Smart-LLaMA: two-stage post-training of large language models for smart contract vulnerability detection and explanation}.
\newblock \bibinfo{journal}{\emph{arXiv preprint arXiv:2411.06221}} (\bibinfo{year}{2024}).
\newblock


\bibitem[Yu et~al\mbox{.}(2025a)]%
        {yu2025sael}
\bibfield{author}{\bibinfo{person}{Lei Yu}, \bibinfo{person}{Shiqi Cheng}, \bibinfo{person}{Zhirong Huang}, \bibinfo{person}{Jingyuan Zhang}, \bibinfo{person}{Chenjie Shen}, \bibinfo{person}{Junyi Lu}, \bibinfo{person}{Li Yang}, \bibinfo{person}{Fengjun Zhang}, {and} \bibinfo{person}{Jiajia Ma}.} \bibinfo{year}{2025}\natexlab{a}.
\newblock \showarticletitle{Sael: Leveraging large language models with adaptive mixture-of-experts for smart contract vulnerability detection}.
\newblock \bibinfo{journal}{\emph{arXiv preprint arXiv:2507.22371}} (\bibinfo{year}{2025}).
\newblock


\bibitem[Yu et~al\mbox{.}(2025b)]%
        {yu2025smart}
\bibfield{author}{\bibinfo{person}{Lei Yu}, \bibinfo{person}{Zhirong Huang}, \bibinfo{person}{Hang Yuan}, \bibinfo{person}{Shiqi Cheng}, \bibinfo{person}{Li Yang}, \bibinfo{person}{Fengjun Zhang}, \bibinfo{person}{Chenjie Shen}, \bibinfo{person}{Jiajia Ma}, \bibinfo{person}{Jingyuan Zhang}, \bibinfo{person}{Junyi Lu}, {et~al\mbox{.}}} \bibinfo{year}{2025}\natexlab{b}.
\newblock \showarticletitle{Smart-LLaMA-DPO: Reinforced Large Language Model for Explainable Smart Contract Vulnerability Detection}.
\newblock \bibinfo{journal}{\emph{Proceedings of the ACM on Software Engineering}} \bibinfo{volume}{2}, \bibinfo{number}{ISSTA} (\bibinfo{year}{2025}), \bibinfo{pages}{182--205}.
\newblock


\bibitem[Yu et~al\mbox{.}(2023a)]%
        {yu2023pscvfinder}
\bibfield{author}{\bibinfo{person}{Lei Yu}, \bibinfo{person}{Junyi Lu}, \bibinfo{person}{Xianglong Liu}, \bibinfo{person}{Li Yang}, \bibinfo{person}{Fengjun Zhang}, {and} \bibinfo{person}{Jiajia Ma}.} \bibinfo{year}{2023}\natexlab{a}.
\newblock \showarticletitle{PSCVFinder: A Prompt-Tuning Based Framework for Smart Contract Vulnerability Detection}. In \bibinfo{booktitle}{\emph{2023 IEEE 34th International Symposium on Software Reliability Engineering (ISSRE)}}. IEEE, \bibinfo{pages}{556--567}.
\newblock


\bibitem[Yu et~al\mbox{.}(2023b)]%
        {yu2023money}
\bibfield{author}{\bibinfo{person}{Lei Yu}, \bibinfo{person}{Fengjun Zhang}, \bibinfo{person}{Jiajia Ma}, \bibinfo{person}{Li Yang}, \bibinfo{person}{Yuanzhe Yang}, {and} \bibinfo{person}{Wei Jia}.} \bibinfo{year}{2023}\natexlab{b}.
\newblock \showarticletitle{Who Are the Money Launderers? Money Laundering Detection on Blockchain via Mutual Learning-Based Graph Neural Network}. In \bibinfo{booktitle}{\emph{2023 International Joint Conference on Neural Networks (IJCNN)}}. IEEE, \bibinfo{pages}{1--8}.
\newblock


\bibitem[Zheng et~al\mbox{.}(2024)]%
        {zheng2024llamafactory}
\bibfield{author}{\bibinfo{person}{Yaowei Zheng}, \bibinfo{person}{Richong Zhang}, \bibinfo{person}{Junhao Zhang}, \bibinfo{person}{Yanhan Ye}, {and} \bibinfo{person}{Zheyan Luo}.} \bibinfo{year}{2024}\natexlab{}.
\newblock \showarticletitle{Llamafactory: Unified efficient fine-tuning of 100+ language models}.
\newblock \bibinfo{journal}{\emph{arXiv preprint arXiv:2403.13372}} (\bibinfo{year}{2024}).
\newblock


\bibitem[Zhuang et~al\mbox{.}(2020)]%
        {zhuang2020smart}
\bibfield{author}{\bibinfo{person}{Yuan Zhuang}, \bibinfo{person}{Zhenguang Liu}, \bibinfo{person}{Peng Qian}, \bibinfo{person}{Qi Liu}, \bibinfo{person}{Xiang Wang}, {and} \bibinfo{person}{Qinming He}.} \bibinfo{year}{2020}\natexlab{}.
\newblock \showarticletitle{Smart Contract Vulnerability Detection using Graph Neural Network.}. In \bibinfo{booktitle}{\emph{IJCAI}}. \bibinfo{pages}{3283--3290}.
\newblock


\bibitem[Zou et~al\mbox{.}(2019)]%
        {zou2019smart}
\bibfield{author}{\bibinfo{person}{Weiqin Zou}, \bibinfo{person}{David Lo}, \bibinfo{person}{Pavneet~Singh Kochhar}, \bibinfo{person}{Xuan-Bach~Dinh Le}, \bibinfo{person}{Xin Xia}, \bibinfo{person}{Yang Feng}, \bibinfo{person}{Zhenyu Chen}, {and} \bibinfo{person}{Baowen Xu}.} \bibinfo{year}{2019}\natexlab{}.
\newblock \showarticletitle{Smart contract development: Challenges and opportunities}.
\newblock \bibinfo{journal}{\emph{IEEE Transactions on Software Engineering}} \bibinfo{volume}{47}, \bibinfo{number}{10} (\bibinfo{year}{2019}), \bibinfo{pages}{2084--2106}.
\newblock


\end{thebibliography}
\end{document}